\documentclass[twocolumn,english,prb,showpacs,preprintnumbers,amsmath,amssymb]{revtex4}
\usepackage[T1]{fontenc}
\usepackage[latin1]{inputenc}
\usepackage{bm}
\usepackage{amsmath}
\usepackage{graphicx}
\usepackage{amssymb}
\usepackage{esint}

\makeatletter
\@ifundefined{textcolor}{}
{%
 \definecolor{BLACK}{gray}{0}
 \definecolor{WHITE}{gray}{1}
 \definecolor{RED}{rgb}{1,0,0}
 \definecolor{GREEN}{rgb}{0,1,0}
 \definecolor{BLUE}{rgb}{0,0,1}
 \definecolor{CYAN}{cmyk}{1,0,0,0}
 \definecolor{MAGENTA}{cmyk}{0,1,0,0}
 \definecolor{YELLOW}{cmyk}{0,0,1,0}
 }

\usepackage{epsfig}

\usepackage{dcolumn}% Align table columns on decimal point
\usepackage{bm}% bold math
\usepackage{wasysym}

\usepackage{marvosym}

\@ifundefined{definecolor}
 {\usepackage{color}}{}

\usepackage{hyperref}

\makeatletter

\usepackage{ulem}

\makeatother

\makeatother

\usepackage{babel}

\makeatother

\usepackage{babel}

\begin{document}

\title{Temperature dependent resistivity in bilayer graphene due to flexural
phonons}
\author{H. Ochoa,$^1$ Eduardo V. Castro,$^{1,2}$ M. I. Katsnelson,$^3$ and F. Guinea$^1$}
{\affiliation{$^1$Instituto de Ciencia de Materiales de
Madrid (CSIC), Sor Juana In\'es de la Cruz 3, E-28049 Madrid,
Spain. \\ 
$^2$Centro de F\'isica do Porto, Rua do Campo Alegre 687, P-4169-007 
  Porto, Portugal\\
$^3$Radboud University Nijmegen, Institute for Molecules and Materials,
NL-6525 AJ Nijmegen, The Netherlands}

\begin{abstract}
We have studied electron scattering by out-of-plane (flexural) phonons
in doped suspended bilayer graphene. We have found the bilayer membrane
to follow the qualitative behavior of the monolayer cousin. In the
bilayer, different electronic structure combine with different electron-phonon
coupling to give the same parametric dependence in resistivity, and
in particular the same temperature $T$ behavior. In parallel with
the single layer, flexural phonons dominate the phonon contribution
to resistivity in the absence of strain, where a density independent
mobility is obtained. This contribution is strongly suppressed by
tension, and in-plane phonons become the dominant contribution in
strained samples. Among
the quantitative differences an important one has been identified:
room $T$ mobility in bilayer graphene is substantially higher than
in monolayer. The origin of quantitative differences has been unveiled.
\end{abstract}


\maketitle
%-----------------------------------------------------------------------------%
%-----------------------------------------------------------------------------%

\section{Introduction}

Bilayer graphene continues to attract a great deal of attention because
of both fascinating fundamental physics\cite{FMY09} and possible
applications.\cite{ZTG+09} Recent realization of suspended monolayer\cite{BSJ+08,DSB+08}
and bilayer\cite{FMY09} graphene samples made possible a direct probe
of the intrinsic, unusual properties of these systems. In particular,
intrinsic scattering mechanisms limiting mobility may now be unveiled.\cite{MNK+07}
It has been recently shown that in suspended, non-strained monolayer
graphene room temperature $T$ mobility is limited to values observed
for samples on substrate due to scattering by out of plane --- \emph{flexural}
--- acoustic phonons.\cite{COK+10} This limitation can, however,
be avoided by applying tension. Bilayer graphene has a different low
energy electronic behavior as well as different electron-phonon coupling.
It is then natural to wonder what is the situation in the bilayer
regarding electron scattering by acoustic phonons, and in particular
by flexural phonons (FPs).

In the present work the $T$ dependent resistivity due to scattering
by both acoustic in-plane phonons and FPs in doped, suspended bilayer
graphene, has been investigated. We have found the bilayer membrane
to follow the qualitative behavior of the monolayer parent.\cite{MvO10,COK+10}
Explicitly, at experimentally relevant $T$, the non-strained samples
show quadratic in $T$ resistivity with logarithmic correction, $\varrho\sim T^{2}\ln(T)$,
and constant mobility. Electron scattering by two FPs gives the main
contribution to the resistivity in this case, and is responsible for
the $T^{2}$ dependence. Suspended samples may also be under strain
either due to the charging gate\cite{OCK+10} or due to the experimental
procedure to get suspended samples, or even by applying strain in
a controlled way.\cite{BMC+09,CRB+09} Under uniaxial or isotropic
strain $u$ the $T$ dependence of resistivity due to FPs becomes
quartic at high strain $u\gg u^{*}$, $\varrho\sim T^{4}/u^{3}$,
and quadratic at low strain $u\ll u^{*}$, $\varrho\sim T^{2}/u$,
where $u^{*}\approx10^{-4}T(\mbox{K})$. These contributions are weaker
than that coming from scattering by in-plane phonons, and in strained samples
the latter dominates resistivity, as has been found for monolayer
graphene.\cite{COK+10} An interesting quantitative difference with
respect to suspended monolayer has been found. In the latter, room
$T$ mobility $\mu$ is limited to values obtained for samples on
substrate due to FPs, $\mu\sim1\,\,\mbox{m}^{2}/(\mbox{Vs})$.\cite{COK+10}
In bilayer, quantitative differences in electron-phonon coupling and
elastic constants lead to room $T$ enhanced $\mu\sim10-20\,\,\mbox{m}^{2}/(\mbox{Vs})$,
even in non-strained samples.

The paper is organized as follows. In Sec.~\ref{sec:phonons} we introduce long wavelength acoustic phonons in the framework of elasticity theory. We show how the dispersion relation of FPs are affected by the presence of tension over the sample. Then, we review the electronic low-energy description of bilayer graphene and deduce the electron-phonon coupling within this approach in Sec.~\ref{sec:low+coupling}. The variational approach used in order to study the $T$ dependent resistivity due to scattering by in-plane and FPs and a summary of our results in different regimes of $T$ and strain are presented in Sec.~\ref{sec:Tres}. Sec.~\ref{sec:Discuss} is devoted to discuss the implications of these results, the differences between monolayer and bilayer and some experimental consequences. Finally, we expose our conclusions in Sec.~\ref{sec:Conclusions}. Some technical aspects are treated in detail in appendices. In Appendix.~\ref{sec:APPcollInt} we present the collision integral due to scattering by acoustic phonons, and in Appendix~\ref{sec:APPlinCollInt} its linearized form is derived. Details on the calculation of the resistivity using the variational method are given in Appendix~\ref{sec:APPres}. In Appendix~\ref{secapp:anharmonic} we discuss how anharmonic effects are partially suppressed by the presence of strain.

%-----------------------------------------------------------------------------%
%-----------------------------------------------------------------------------%

\section{Acoustic phonons}

\label{sec:phonons}

At long-wavelengths the elastic behavior of monolayer graphene is
well approximated by that of an isotropic continuum membrane\cite{NGPrmp,ZKF09}
whose free-energy reads,\citep{LL7,NelsonSMMemb}\begin{equation}
\mathcal{F}=\frac{1}{2}\kappa\int dxdy(\nabla^{2}h)^{2}+\frac{1}{2}\int dxdy(\lambda u_{ii}^{2}+2\mu u_{ij}^{2}).\label{eq:fe}\end{equation}
The first and second terms in Eq.~\eqref{eq:fe} represent the bending
and stretching energies, respectively. Summation over indices is assumed.
In-plane distortions are denoted by $\mathbf{u}(\mathbf{r})$ and
out-of-plane $h(\mathbf{r})$, with $\mathbf{r}=(x,y)$, such that
the new position is $\vec{X}(\mathbf{r})=(x,y,0)+[u_{x}(\mathbf{r}),u_{y}(\mathbf{r}),h(\mathbf{r})]$.
To lowest order in gradients of the deformations the strain tensor
appearing in Eq.~\eqref{eq:fe} is\begin{equation}
u_{ij}=\frac{1}{2}[\partial_{i}u_{j}+\partial_{j}u_{i}+(\partial_{i}h)(\partial_{j}h)],\label{eq:uij}\end{equation}
 and owing to the same argument the factor $\sqrt{g}=\sqrt{1+|\nabla h|^{2}}$
in the measure is neglected. The parameter $\kappa$ is the bending
rigidity and $\lambda$ and $\mu$ are in-plane elastic constants.
Typical parameters for graphene are $\kappa\approx1\,\mbox{eV}$ and
$\mu\simeq3\lambda\approx9\,\mbox{eV}\textrm{\AA}^{-2}$,\cite{KSY01,LWK+08,ZKF09}
with mass density $\rho=7.6\times10^{-7}\,\mbox{kg/m}^{2}$.

In the case of bilayer graphene, as long as we are not at too high
$T$ to excite optical phonons we may, from the elastic point of view,
regard bilayer graphene as a thick membrane with mass density and
elastic constants twice as high as those for single layer.\cite{ZLK+10}

The dynamics of the displacement fields is here studied in the harmonic
approximation by introducing the Fourier series $\mathbf{u}(\mathbf{r})=\mathcal{V}^{-\frac{1}{2}}\sum_{\mathbf{q}}\mathbf{u}_{\mathbf{q}}e^{i\mathbf{q}.\mathbf{r}}$
and $h(\mathbf{r})=\mathcal{V}^{-\frac{1}{2}}\sum_{\mathbf{q}}h_{\mathbf{q}}e^{i\mathbf{q}.\mathbf{r}}$,
where $\mathcal{V}$ is the volume of the system.

\subsection{In-plane phonons}

The decoupled in-plane phonon modes are obtained in the usual way
by changing to longitudinal $u_{\mathbf{q}}^{L}=\mathbf{u}_{\mathbf{q}}\cdot\mathbf{q}/q$
and transverse $u_{\mathbf{q}}^{T}=\mathbf{u}_{\mathbf{q}}\cdot(\hat{e}_{z}\times\mathbf{q}/q)$
displacement fields. The dispersion relations have the usual linear
behavior in momentum and are given by\begin{eqnarray}
\omega_{\mathbf{q}}^{L} & = & v_{L}q,\nonumber \\
\omega_{\mathbf{q}}^{T} & = & v_{T}q,\label{eq:inpDisp}\end{eqnarray}
 with $v_{L}=\sqrt{\frac{2\mu+\lambda}{\rho}}$ and $v_{T}=\sqrt{\frac{\mu}{\rho}}$.
Typical values for monolayer and bilayer graphene are $v_{L}\simeq2.1\times10^{4}\,\mbox{m}/\mbox{s}$
and $v_{T}\simeq1.4\times10^{4}\,\mbox{m}/\mbox{s}$.

\subsection{Flexural phonons}

\subsubsection{Non-strained case}

The quadratic behavior in out-of-plane displacements of the strain
tensor in Eq.~\eqref{eq:uij} implies that FP modes are driven by
the bending rigidity term. The resulting FP dispersion relation is
quadratic,\citep{MO08,NGPrmp}\begin{equation}
\omega_{\mathbf{q}}^{F}=\alpha q^{2},\label{eq:flexDispUnStr}\end{equation}
 with $\alpha=\sqrt{\frac{\kappa}{\rho}}$. The typical value is $\alpha\simeq4.6\times10^{-7}\,\mbox{m}^{2}/\mbox{s}$.

\subsubsection{Strained case}

Suspended samples may be under tension either due to the load imposed
by the back gate or as a result of the fabrication process, or both.
The case of a clamped graphene membrane hanging over a tranche of
size $L$, relevant for conventional two-contacts measurements in
suspended samples, has been considered in Ref.~\onlinecite{FGK08}.

Once the membrane is under tension a static deformation configuration
is expected at equilibrium. The phonon modes may be obtained by assuming
that both in-plane $\mathbf{u}(\mathbf{r})$ and flexural $h(\mathbf{r})$
fields have dynamic components which add to their static background:
$\mathbf{u}(\mathbf{r})=\mathbf{u}_{st}(\mathbf{r})+\mathbf{u}_{dyn}(\mathbf{r})$
and $h(\mathbf{r})=h_{st}(\mathbf{r})+h_{dyn}(\mathbf{r})$. For the
case of the clamped membrane considered in Ref.~\onlinecite{FGK08}
we have $\mathbf{u}_{st}(\mathbf{r})=[u_{x,st}(x),0]$ with $u_{x,st}(x)$
a linear function of $x$, while $h_{st}(x)$ may be approximated
by a parabola.

Let us consider the general static displacement vector field $\mathbf{d}_{st}(\mathbf{r})=\left[u_{x,st}(\mathbf{r}),u_{y,st}(\mathbf{r}),h_{st}(\mathbf{r})\right]$
and the associated strain tensor $u_{ij,st}(\mathbf{r})=\frac{1}{2}(\partial_{i}u_{j,st}+\partial_{j}u_{i,st}+\partial_{i}h_{st}\partial_{j}h_{st})$.
In-plane phonons are not affected by the static component but the
FP dispersion changes considerably. This is a consequence of new harmonic
terms appearing due to coupling between in-plane static deformation
and out-of-plane vibrations in the full strain tensor in Eq.~\eqref{eq:uij}.
The resulting FP dispersion relation may be obtained using a \emph{local
approximation} expected to hold for $L\gg\ell=v_{F}\tau\gg k_{F}^{-1}$,
where $v_{F}$ is the Fermi velocity, $k_{F}$ the Fermi momentum,
and $\tau$ a characteristic collision time. The result reads
\begin{equation}
\omega_{\mathbf{q}}^{F}(\mathbf{r})=q\sqrt{\frac{\kappa}{\rho}q^{2}+u_{ii,st}(\mathbf{r})\frac{\lambda}{\rho}+u_{ij,st}(\mathbf{r})\frac{2\mu}{\rho}\frac{q_{i}q_{j}}{q^{2}}}.\label{eq:flexDispStr1}\end{equation}
 For the particular case of isotropic strain where $u_{xx}=u_{yy}$
and $u_{xy}=0$ the dispersion relation can be cast in the form\begin{equation}
\omega_{\mathbf{q}}^{F}(\mathbf{r})=q\sqrt{\frac{\kappa}{\rho}q^{2}+u_{ii,st}(\mathbf{r})\frac{\lambda+\mu}{\rho}}.\label{eq:flexDispStr2}\end{equation}

Here we will give particular emphasis to the clamped membrane case
where the FP dispersion is given by\begin{equation}
\omega_{\mathbf{q}}^{F}=q\sqrt{\frac{\kappa}{\rho}q^{2}+\bar{u}\frac{\lambda+2\mu}{\rho}-\bar{u}\frac{2\mu}{\rho}\sin^{2}\phi_{\mathbf{q}}},\label{eq:flexDispStr3}\end{equation}
with $\bar{u}\equiv u_{xx}$ and $\phi_{\mathbf{q}}=\arctan(q_{y}/q_{x})$.
In order to keep the problem within analytical treatment we will use
an effective isotropic dispersion relation, obtained by dropping the
angular dependence contribution,\begin{equation}
\omega_{\mathbf{q}}^{F}\simeq q\sqrt{\alpha^{2}q^{2}+\bar{u}v_{L}^{2}}.\label{eq:flexDispStr4}\end{equation}
 Since we are mainly interested in transport such an approximation
has the advantage that backward scattering is still correctly accounted
for.

%-----------------------------------------------------------------------------%
%-----------------------------------------------------------------------------%

\section{Electron-phonon interaction}

\label{sec:low+coupling}

\subsection{Low energy description for bilayer graphene }

At low energies the 2-band effective model provides a good approximate
description for $\pi-$electrons in bilayer graphene.\cite{NGPrmp}
The $2\times2$ Hamiltonian can be cast in the form $\mathcal{H}_{eff}=\sum_{\mathbf{k}}\psi_{\mathbf{k}}\mathcal{H}_{\mathbf{k}}\psi_{\mathbf{k}}$,
with \begin{equation}
\mathcal{H}_{\mathbf{k}}=\frac{\hbar^{2}}{2m}\left(\begin{array}{cc}
0 & \left(k_{x}-ik_{y}\right)^{2}\\
\left(k_{x}+ik_{y}\right)^{2} & 0\end{array}\right),\label{eq:Hdirac_bilayer}\end{equation}
where the two component spinor $\psi_{\mathbf{k}}^{\dagger}=[a_{\mathbf{k}}^{\dagger},b_{\mathbf{k}}^{\dagger}]$
stems from the two sublattices not connected by the interlayer hopping,
$t_{\perp}\approx0.3\,\mbox{eV}$. The coupling $t_{\perp}$ between
layers sets the effective mass $2m=t_{\perp}/v_{F}^{2}$, with $v_{F}\approx10^{6}\,\mbox{m}/\mbox{s}$
the Fermi velocity in monolayer graphene. Equation~\eqref{eq:Hdirac_bilayer}
is valid at valley $K$, at valley $K'$ we have $\mathcal{H}_{\mathbf{k}}\rightarrow\mathcal{H}_{\mathbf{k}}^{T}$.
Here we are interested in electron scattering processes induced by
emission or absorption of long wavelength acoustic phonons and hence
intervalley scattering is not allowed. Thus we may concentrate on
one valley only. The Hamiltonian can be diagonalized introducing the
rotated operators \begin{equation}
d_{\mathbf{k}}=\frac{1}{\sqrt{2}}\left(\begin{array}{cc}
e^{i\theta_{\mathbf{k}}} & e^{-i\theta_{\mathbf{k}}}\\
e^{i\theta_{\mathbf{k}}} & -e^{-i\theta_{\mathbf{k}}}\end{array}\right)\psi_{\mathbf{k}},\label{eq:rot}\end{equation}
where $\theta_{\mathbf{k}}=\arctan(k_{y}/k_{x})$, with $d_{\mathbf{k}}^{\dagger}=[e_{\mathbf{k}}^{\dagger},h_{\mathbf{k}}^{\dagger}]$
defined such that $e_{\mathbf{k}}^{\dagger}$ stands for electron-like
(positive energy) excitations and $h_{\mathbf{k}}^{\dagger}$ for
hole-like (negative energy) excitations, from which we get\begin{equation}
\mathcal{H}=\sum_{\mathbf{k}}\varepsilon\left(\mathbf{k}\right)[e_{\mathbf{k}}^{\dagger}e_{\mathbf{k}}-h_{\mathbf{k}}^{\dagger}h_{\mathbf{k}}],\label{eq:Heh}\end{equation}
 with $\varepsilon\left(\mathbf{k}\right)=\hbar^{2}k^{2}/(2m)$.

Along the paper we will compare the results obtained for bilayer graphene
with those valid in monolayer. The latter is described using the effective
Dirac-like Hamiltonian,\cite{NGPrmp} which holds in the low energy
sector we are interested here.

\subsection{Coupling between electrons and phonons}

\label{sub:coupling}

The coupling between electrons and the vibrations of the underlying
lattice either in bilayer or monolayer graphene has two main sources.
Long wavelength acoustic phonons induce an effective local potential
called deformation potential and proportional to the local contraction
or dilation of the lattice,\citep{NGPrmp}\[
V_{1}(\mathbf{r})=g_{0}u_{ii}(\mathbf{r}),\]
where $g_{0}$ is the bare deformation potential constant, whose value
is in the range $g_{0}\approx20-30\,\mbox{eV}$.\cite{SA02a} The
respective interaction Hamiltonian is diagonal in sublattice indices
and reads \begin{equation}
\mathcal{H}_{1}=\sum_{\mathbf{k}}\psi_{\mathbf{k}}^{\dagger}\left[V_{1}(\mathbf{k},\mathbf{k}')\mathbb{I}\right]\psi_{\mathbf{k}'},\label{eq:H1}\end{equation}
 where $\mathbb{I}$ is the $2\times2$ identity matrix and $V_{1}(\mathbf{k},\mathbf{k}')$
is the Fourier transform of the deformation potential\begin{equation}
V_{1}(\mathbf{k},\mathbf{k}')=\mathcal{V}^{-1}g_{0}\int d\mathbf{r}e^{i(\mathbf{k}'-\mathbf{k})\cdot\mathbf{r}}u_{ii}\left(\mathbf{r}\right).\label{eq:V1}\end{equation}
 Equation~\eqref{eq:H1} is valid both for monolayer and bilayer
graphene. Since we are interested in doped systems we take into account
screening by substituting $V_{1}(\mathbf{k},\mathbf{k}')$ with $V_{1}\left(\mathbf{k},\mathbf{k}'\right)/\epsilon\left(\mathbf{k}-\mathbf{k}'\right)$,
where we take a Thomas-Fermi like dielectric function\begin{equation}
\epsilon\left(\mathbf{q}\right)=1+\frac{e^{2}\mathcal{D}\left(E_{F}\right)}{2\epsilon_{0}q},\end{equation}
and $\mathcal{D}\left(E_{F}\right)$ is the density of states at the
Fermi energy, which is given by $\mathcal{D}\left(E_{F}\right)=\frac{2E_{F}}{\pi\hbar^{2}v_{F}^{2}}=\frac{2k_{F}}{\pi\hbar v_{F}}$
in the case of monolayer graphene and by $\mathcal{D}\left(E_{F}\right)=\frac{t_{\perp}}{\pi\hbar^{2}v_{F}^{2}}$
in the case of bilayer.

It is convenient to define $g\equiv g_{0}/\epsilon(k_{F})$ for single
layer graphene, which gives a density independent screened deformation
potential \begin{equation}
g\approx\frac{g_{0}}{e^{2}/(\pi\epsilon_{0}\hbar v_{F})}\approx\frac{g_{0}}{8.75}\approx2-3.5\,\,\mbox{eV}.\label{eq:g}\end{equation}
Note that the value just obtained is in complete agreement with recent
\emph{ab initio} calculations which give $g\approx3\,\mbox{eV}$.\cite{CJS10}
It will become clear in Sec.~\ref{sub:ResContribInP} that $g$ as
defined in Eq.~\eqref{eq:g} is the relevant deformation potential
electron-phonon parameter in single layer graphene. For bilayer graphene
$g(q)=g_{0}/\epsilon(q)$ gives\begin{equation}
g(k_{F})\approx\frac{g_{0}}{e^{2}t_{\perp}/(2\pi\epsilon_{0}\hbar^{2}v_{F}^{2}k_{F})}\approx\frac{g_{0}}{11.25}\sqrt{n}\approx(2-3)\sqrt{n}\,\,\mbox{eV},\label{eq:gBil}\end{equation}
with $n$ in $10^{12}\,\mbox{cm}^{-2}$. We may then write a $q$
dependent deformation potential electron-phonon parameter which has
the form $g_{M}(q)=gq/k_{F}$ for monolayer graphene, and $g_{B}(q)=g2\hbar v_{F}q/t_{\perp}$
for bilayer.

Phonons can also couple to electrons in monolayer and bilayer graphene
by changes in bond length and bond angle between carbon atoms. In
this case the electron-phonon interaction can be written as due to
an effective gauge field,\citep{MNK+06,KNssc07,KG08,GHD08}\[
e\mathbf{A}_{elastic}=\frac{\beta}{a}\Bigl[\begin{array}{cc}
\frac{1}{2}(u_{xx}-u_{yy}), & -u_{xy}\end{array}\Bigr],\]
where $\beta\approx-\partial\log t/\partial\log a\sim2-3$,\citep{SA02a}
with $t$ the in-plane nearest neighbor hopping parameter and $a$
the carbon-carbon distance ($t_{0}\approx3\,\mbox{eV}$ and $a_{0}\approx1.4\,\textrm{\AA}$).
In the case of bilayer graphene the resulting interaction Hamiltonian
is obtained by introducing the gauge potential into Eq~\eqref{eq:Hdirac_bilayer},
following the minimal coupling prescription, and keeping only first
order terms in electron-phonon coupling. Then we arrive at \begin{equation}
\mathcal{H}_{2}=\frac{\hbar}{2m}\sum_{\mathbf{k},\mathbf{k}'}\psi_{\mathbf{k}}^{\dagger}\left[\begin{array}{cc}
0 & \left(\pi_{\mathbf{k}}^{-}+\pi_{\mathbf{k}'}^{-}\right)A_{\mathbf{k},\mathbf{k}'}^{-}\\
\left(\pi_{\mathbf{k}}^{+}+\pi_{\mathbf{k}'}^{+}\right)A_{\mathbf{k},\mathbf{k}'}^{+} & 0\end{array}\right]\psi_{\mathbf{k}'},\label{eq:H2_bi}\end{equation}
 where $\pi_{\mathbf{k}}^{\pm}=ke^{\pm i\theta_{\mathbf{k}}}$, $A_{\mathbf{k},\mathbf{k}'}^{\pm}=V_{2,x}(\mathbf{k},\mathbf{k}')\pm iV_{2,y}(\mathbf{k},\mathbf{k}')$
and the vector $\mathbf{V}_{2}=(V_{2,x},V_{2,y})$ is defined as \begin{align}
V_{2,x}(\mathbf{k},\mathbf{k}')= & \frac{\hbar\beta}{a}\mathcal{V}^{-1}\int d\mathbf{r}e^{i(\mathbf{k}'-\mathbf{k})\cdot\mathbf{r}}\frac{1}{2}\left[u_{xx}(\mathbf{r})-u_{yy}(\mathbf{r})\right],\nonumber \\
V_{2,y}(\mathbf{k},\mathbf{k}')= & -\frac{\hbar\beta}{a}\mathcal{V}^{-1}\int d\mathbf{r}e^{i(\mathbf{k}'-\mathbf{k})\cdot\mathbf{r}}u_{xy}(\mathbf{r}).\label{eq:V2}\end{align}
An estimate of the electron-phonon coupling strength due to $V_{2}$
is given by $k_{F}\hbar^{2}\beta/(ma)\approx(8-12)\sqrt{n}\,\mbox{eV}$,
with $n$ in $10^{12}\,\mbox{cm}^{-2}$. In the case of single layer
graphene the resulting interaction Hamiltonian reads\begin{equation}
\mathcal{H}_{2}=v_{F}\sum_{\mathbf{k},\mathbf{k}'}\psi_{\mathbf{k}}^{\dagger}\bm{\sigma}\cdot\mathbf{V}_{2}(\mathbf{k},\mathbf{k}')\psi_{\mathbf{k}'},\label{eq:H2_mono}\end{equation}
where $\bm{\sigma}=(\sigma_{x},\sigma_{y})$ is the vector of Pauli
matrices, and the two component spinor $\psi_{\mathbf{k}}^{\dagger}=[a_{\mathbf{k}}^{\dagger},b_{\mathbf{k}}^{\dagger}]$
is reminiscent of the two sublattices of the honeycomb lattice. An
estimate of the respective electron-phonon coupling strength is given
by $v_{F}\hbar\beta/a\approx10-15\,\mbox{eV}$.

The electron-phonon interaction Hamiltonian is the sum of the two
terms shown above, $H_{ep}=\mathcal{H}_{1}+\mathcal{H}_{2}$. Phonons
enter through the strain tensor $u_{ij}$ which we have seen can be
written in terms of static and dynamic components; the static ones
being zero for zero load. There are purely static terms which do not
contribute to electron-phonon scattering and will be dropped (see
Ref.~\onlinecite{FGK08}). Quantizing the dynamic part of the displacement
fields,\citep{ASqftCondMatt} and introducing usual destruction and
creation operators $a_{\mathbf{q}}^{\nu}$ and $(a_{\mathbf{q}}^{\nu})^{\dagger}$
for in-plane phonons $\mathbf{q}$ and polarization
$\nu=L,T$, we can write the $\mathbf{q}$ component of the in-plane
displacement as,\begin{equation}
u_{\mathbf{q}}^{L/T}=\sqrt{\frac{\hbar}{2\rho\omega_{\mathbf{q}}^{\nu}}}\left[a_{\mathbf{q}}^{L/T}+(a_{-\mathbf{q}}^{L/T})^{\dagger}\right].\label{eq:phononOpInp}\end{equation}
For FPs we introduce the bosonic fields
$a_{\mathbf{q}}^{F}$ and $(a_{\mathbf{q}}^{F})^{\dagger}$, and write
the $\mathbf{q}$ component of the out-of-plane displacement as,\begin{equation}
h_{\mathbf{q}}=\sqrt{\frac{\hbar}{2\rho\omega_{\mathbf{q}}^{F}}}\left[a_{\mathbf{q}}^{F}+(a_{-\mathbf{q}}^{F})^{\dagger}\right].\label{eq:phononOpF}\end{equation}
The electron-phonon interaction Hamiltonian may then be written either
in monolayer or bilayer graphene as,\begin{widetext}\begin{multline}
H_{ep}=\sum_{\mathbf{k},\mathbf{k}'}\left(a_{\mathbf{k}}^{\dagger}a_{\mathbf{k}'}+b_{\mathbf{k}}^{\dagger}b_{\mathbf{k}'}\right)\left\{ \sum_{\nu,\mathbf{q}}V_{1,\mathbf{q}}^{\nu}\left[a_{\mathbf{q}}^{\nu}+(a_{-\mathbf{q}}^{\nu})^{\dagger}\right]\delta_{\mathbf{k}',\mathbf{k}-\mathbf{q}}+\sum_{\mathbf{q},\mathbf{q}'}V_{1,\mathbf{q},\mathbf{q}'}^{F}\left[a_{\mathbf{q}}^{F}+(a_{-\mathbf{q}}^{F})^{\dagger}\right]\left[a_{\mathbf{q}'}^{F}+(a_{-\mathbf{q}'}^{F})^{\dagger}\right]\delta_{\mathbf{k}',\mathbf{k}-\mathbf{q}-\mathbf{q}'}\right\} \\
+\sum_{\mathbf{k},\mathbf{k}'}\left\{ \sum_{\nu,\mathbf{q}}V_{2,\mathbf{q}}^{\nu}a_{\mathbf{k}}^{\dagger}b_{\mathbf{k}'}\left[a_{\mathbf{q}}^{\nu}+(a_{-\mathbf{q}}^{\nu})^{\dagger}\right]\delta_{\mathbf{k}',\mathbf{k}-\mathbf{q}}+\sum_{\mathbf{q},\mathbf{q}'}V_{2,\mathbf{q},\mathbf{q}'}^{F}a_{\mathbf{k}}^{\dagger}b_{\mathbf{k}'}\left[a_{\mathbf{q}}^{F}+(a_{-\mathbf{q}}^{F})^{\dagger}\right]\left[a_{\mathbf{q}'}^{F}+(a_{-\mathbf{q}'}^{F})^{\dagger}\right]\delta_{\mathbf{k}',\mathbf{k}-\mathbf{q}-\mathbf{q}'}+\mbox{h.c.}\right\} .\label{eq:HepCompact}\end{multline}
\end{widetext} For monolayer graphene the matrix elements read,\begin{eqnarray}
V_{1,\mathbf{q}}^{L} & = & \frac{g_{0}}{\epsilon(q)}iq\sqrt{\frac{\hbar}{2\mathcal{V}\rho\omega_{\mathbf{q}}^{L}}},\nonumber \\
V_{1,\mathbf{q},\mathbf{q}'}^{F} & = & -\frac{g_{0}}{\epsilon(|\mathbf{q}+\mathbf{q}'|)}qq'\cos(\phi_{\mathbf{q}}-\phi_{\mathbf{q}'})\frac{\hbar}{4\mathcal{V}\rho\sqrt{\omega_{\mathbf{q}}^{F}\omega_{\mathbf{q}'}^{F}}},\nonumber \\
V_{1,\mathbf{q}}^{F} & = & \frac{g_{0}}{\epsilon(q)}iq_{i}\partial_{i}h_{st}\sqrt{\frac{\hbar}{2\mathcal{V}\rho\omega_{\mathbf{q}}^{F}}},\nonumber \\
V_{2,\mathbf{q}}^{L} & = & \frac{\hbar v_{F}\beta}{2a}iqe^{i2\phi_{\mathbf{q}}}\sqrt{\frac{\hbar}{2\mathcal{V}\rho\omega_{\mathbf{q}}^{L}}},\nonumber \\
V_{2,\mathbf{q}}^{T} & = & -\frac{\hbar v_{F}\beta}{2a}qe^{i2\phi_{\mathbf{q}}}\sqrt{\frac{\hbar}{2\mathcal{V}\rho\omega_{\mathbf{q}}^{T}}},\nonumber \\
V_{2,\mathbf{q},\mathbf{q}'}^{F} & = & -\frac{\hbar v_{F}\beta}{4a}qq'e^{i(\phi_{\mathbf{q}}-\phi_{\mathbf{q}'})}\frac{\hbar}{2\mathcal{V}\rho\sqrt{\omega_{\mathbf{q}}^{F}\omega_{\mathbf{q}'}^{F}}},\nonumber \\
V_{2,\mathbf{q}}^{F} & = & \frac{\hbar v_{F}\beta}{2a}iq\left[e^{i\phi_{\mathbf{q}}}\partial_{x}h_{st}+e^{-i\phi_{\mathbf{q}}}\partial_{y}h_{st}\right]\sqrt{\frac{\hbar}{2\mathcal{V}\rho\omega_{\mathbf{q}}^{F}}}\,,\label{eq:potential}\end{eqnarray}
with $V_{1,\mathbf{q}}^{T} = 0$ (see also Refs.~\onlinecite{MO08,MvO10}),
and where we have again used
the local approximation. In the case of bilayer graphene only the
matrix elements for the gauge potential change, becoming dependent
on fermionic momenta $\mathbf{k}$ and $\mathbf{k}'$. As can be seen
by comparing Eqs.~\eqref{eq:H2_bi} and~\eqref{eq:H2_mono}, they
take exactly the same form as in Eq.~\eqref{eq:potential} with the
replacement $v_{F}\rightarrow\hbar(\pi_{\mathbf{k}}+\pi_{\mathbf{k}'})/(2m)$.

%-----------------------------------------------------------------------------%
%-----------------------------------------------------------------------------%

\section{Temperature dependent resistivity}

\label{sec:Tres}

Our aim here is to study the $T$ dependent resistivity in suspended
bilayer graphene as a result of the electron-phonon interaction derived
above. We assume the doped regime $E_{F}\gg\hbar/\tau$, where $1/\tau$
is the characteristic electronic scattering rate (due to phonons,
disorder, etc). The doped regime immediately implies $k_{F}^{-1}\ll v_{F}\tau\equiv l$,
where $l$ is the characteristic mean-free path, thus justifying the
use of Boltzmann transport theory (even though graphene's quasiparticles
are chiral the semiclassical approach still holds away from the Dirac
point \citep{AK07,CB09}).

\subsection{The variational approach}

\label{sub:varMethod}

The Boltzmann equation is an integro-deferential equation for the
steady state probability distribution $f_{\mathbf{k}}$.\cite{zimanEP}
It can be generally written as\begin{equation}
\dot{\mathbf{r}}\cdot\nabla_{\mathbf{r}}f_{\mathbf{k}}+\dot{\mathbf{k}}\cdot\nabla_{\mathbf{k}}f_{\mathbf{k}}=\dot{f}_{\mathbf{k}}\bigr|_{\mathrm{scatt}},\label{eq:BoltzEq}\end{equation}
where the terms on the left hand side are due to, respectively, \emph{diffusion}
and \emph{external fields}, while on the right hand side \emph{scattering}
provides the required balance at the steady state. (See Appendix~\ref{sec:APPcollInt}
for an explicit form of $\dot{f}_{\mathbf{k}}\bigr|_{\mathrm{scatt}}$
in the case under study.) The Boltzmann equation is quite intractable in 
practice,
and its linearized version is used instead,\begin{equation}
\dot{\mathbf{r}}\cdot\nabla_{\mathbf{r}}f_{\mathbf{k}}^{(0)}+\dot{\mathbf{k}}\cdot\nabla_{\mathbf{k}}f_{\mathbf{k}}^{(0)}=\delta\dot{f}_{\mathbf{k}}\bigr|_{\mathrm{scatt}},\label{eq:linBoltzEq}\end{equation}
where $\delta\dot{f}_{\mathbf{k}}\bigr|_{\mathrm{scatt}}$ is the
linearized collision integral. (See Appendix~\ref{sec:APPlinCollInt}
for an explicit form of $\delta\dot{f}_{\mathbf{k}}\bigr|_{\mathrm{scatt}}$
in the case under study.) Expanding the distribution probability around
its equilibrium value $f_{\mathbf{k}}^{(0)}=1/\{\exp[(\varepsilon_{\mathbf{k}}-\mu)/k_{B}T]+1\}$,\begin{equation}
f_{\mathbf{k}}=f_{\mathbf{k}}^{(0)}-\frac{\partial f_{\mathbf{k}}^{(0)}}{\partial\varepsilon_{\mathbf{k}}}\Phi_{\mathbf{k}},\label{eq:lin}\end{equation}
and using the equilibrium property that $\dot{f}_{\mathbf{k}}^{(0)}\bigr|_{\mathrm{scatt}}=0$,
it can be seen that $\delta\dot{f}_{\mathbf{k}}\bigr|_{\mathrm{scatt}}$
is linear in $\Phi_{\mathbf{k}}$, and that it can be written as a
linear application in terms of the linear scattering operator $P_{\mathbf{k}}$,
\begin{align}
\delta\dot{f}_{\mathbf{k}}\bigr|_{\mathrm{scatt}} & =P_{\mathbf{k}}\Phi_{\mathbf{k}}\equiv\nonumber \\
 & \equiv-\sum_{\mathbf{k}_{1},\dots,\mathbf{k}_{n}}\mathcal{P}_{\mathbf{k},\mathbf{k}_{1},\dots,\mathbf{k}_{n}}\left(\Phi_{\mathbf{k}}\pm\Phi_{\mathbf{k}_{1}}\dots\pm\Phi_{\mathbf{k}_{n}}\right),\label{eq:linColInt}\end{align}
where $\mathcal{P}_{\mathbf{k},\mathbf{k}_{1},\dots,\mathbf{k}_{n}}$
is a generalized transition rate per unit energy.\cite{zimanEP} (See
Appendix~\ref{sec:APPlinCollInt} for an explicit form of $\mathcal{P}_{\mathbf{k},\mathbf{k}_{1},\dots,\mathbf{k}_{n}}$
in the case under study.) Writing the linearized Boltzmann equation,
Eq.~\eqref{eq:linBoltzEq}, in the form\[
X_{\mathbf{k}}=P_{\mathbf{k}}\Phi_{\mathbf{k}},\]
and defining the inner products,\begin{equation}
\left\langle \Phi,X\right\rangle =\sum_{\mathbf{k}}\Phi_{\mathbf{k}}\left(\dot{\mathbf{r}}\cdot\nabla_{\mathbf{r}}f_{\mathbf{k}}^{(0)}+\dot{\mathbf{k}}\cdot\nabla_{\mathbf{k}}f_{\mathbf{k}}^{(0)}\right),\label{eq:innP1}\end{equation}
and\begin{align}
\left\langle \Phi,P\Phi\right\rangle  & =\sum_{\mathbf{k},\mathbf{k}_{1},\dots,\mathbf{k}_{n}}\Phi_{\mathbf{k}}\mathcal{P}_{\mathbf{k},\mathbf{k}_{1},\dots,\mathbf{k}_{n}}\left(\Phi_{\mathbf{k}}\pm\Phi_{\mathbf{k}_{1}}\dots\pm\Phi_{\mathbf{k}_{n}}\right)\nonumber \\
 & =\frac{1}{(n+1)}\sum_{\mathbf{k},\mathbf{k}_{1},\dots,\mathbf{k}_{n}}\left(\Phi_{\mathbf{k}}\pm\Phi_{\mathbf{k}_{1}}\dots\pm\Phi_{\mathbf{k}_{n}}\right)^{2}\mathcal{P}_{\mathbf{k},\mathbf{k}_{1},\dots,\mathbf{k}_{n}},\label{eq:innP2}\end{align}
the \emph{variational principle} asserts that of all functions $\Phi_{\mathbf{k}}$
satisfying $\left\langle \Phi,X\right\rangle =\left\langle \Phi,P\Phi\right\rangle $,
the solution of the linearized Boltzmann equation gives to the quantity
$\left\langle \Phi,P\Phi\right\rangle /\{\left\langle \Phi,X\right\rangle \}^{2}$
its \emph{minimum} value.\cite{zimanEP} In particular, the resistivity
$\varrho$ can be written as\begin{equation}
\varrho=\frac{1}{g_{d}}\frac{\left\langle \Phi,P\Phi\right\rangle }{\{\left\langle \Phi,X(E=1,\nabla_{\mathbf{r}}f^{(0)}=0)\right\rangle \}^{2}},\label{eq:resistVar}\end{equation}
being thus expected to be a \emph{minimum} for the right solution,\cite{zimanEP}
where $g_{d}$ is the system's degeneracy (for monolayer and bilayer
graphene it is $g_{d}=g_{s}g_{v}=4$ due to spin and valley degeneracies).
The quantity $X(E=1,\nabla_{\mathbf{r}}f^{(0)}=0)$ refers to the
left hand side of Eq.~\eqref{eq:linBoltzEq} in a unit electric field
and no spatial gradients (for example, zero temperature gradient).
It is easy to show that $\left\langle \Phi,X\right\rangle =\mathbf{E}\cdot\mathbf{J}$,
where \[
\mathbf{J}=\sum_{\mathbf{k}}e\mathbf{v}_{\mathbf{k}}\Phi_{\mathbf{k}}\frac{\partial f_{\mathbf{k}}^{(0)}}{\partial\varepsilon_{\mathbf{k}}}\]
is the current per non-degenerate mode (per spin and valley in monolayer
and bilayer graphene). The quantity $\{\left\langle \Phi,X(E=1,\nabla_{\mathbf{r}}f^{(0)}=0)\right\rangle \}^{2}$
is therefore nothing but $\mathcal{V}\mathbf{j}^{2}$, where $\mathbf{j}=\mathbf{J}/\mathcal{V}$
is the current density.

A well known solution to the Boltzmann equation exists when scattering
is elastic, the Fermi surface isotropic, and the transition rate can
be written as $\mathcal{P}_{\mathbf{k},\mathbf{k}'}=\mathcal{P}(k,\theta_{\mathbf{k},\mathbf{k}'})$,
where $\theta_{\mathbf{k},\mathbf{k}'}=\theta_{\mathbf{k}}-\theta_{\mathbf{k}'}$
is the angle between $\mathbf{k}$ and $\mathbf{k}'$.\cite{zimanEP}
Under these conditions the solution reads, \[
\Phi_{\mathbf{k}}=v_{\mathbf{k}}\cdot\left(e\mathbf{E}-\frac{\varepsilon_{k}}{T}\nabla T\right)\tau(k),\]
where $\tau(k)$ is the isotropic scattering rate, and we have written
$\nabla_{\mathbf{r}}f_{\mathbf{k}}^{(0)}=\partial f_{\mathbf{k}}^{(0)}/\partial\varepsilon_{\mathbf{k}}\nabla T$.
Clearly, the later solution for $\Phi_{\mathbf{k}}$ can be cast in
the form $\Phi_{\mathbf{k}}\propto\mathbf{k}\cdot\mathbf{u}$,\cite{foot1,zimanEP} where
$\mathbf{u}$ is a unit vector in the direction of the applied fields.
So, in more complicated cases where there is a departure from the
isotropic conditions and/or from elastic scattering, it is a good
starting point to use Eq.~\eqref{eq:resistVar} with $\Phi_{\mathbf{k}}\propto\mathbf{k}\cdot\mathbf{u}$
to get an approximate (from above) result for the resistivity. Note
that the coefficient multiplying $\mathbf{k}\cdot\mathbf{u}$ is unimportant
as it cancels out. This variational method is equivalent to a linear
response Kubo-Nakano-Mori approach with the perturbation inducing 
scattering treated
in the Born approximation.\cite{IK02}

Here we use the variational method just outlined to get the $T$ dependent
resistivity in bilayer graphene (and monolayer for comparison) due
to scattering by acoustic phonons. In this case, using the quasi-elastic
approximation (see Appendix~\ref{sec:APPlinCollInt}), $\delta\dot{f}_{\mathbf{k}}\bigr|_{\mathrm{scatt}}$
can indeed be cast in the form of Eq.~\eqref{eq:linColInt}, \begin{equation}
\delta\dot{f}_{\mathbf{k}}\bigr|_{\mathrm{scatt}}=-\sum_{\mathbf{k}'}\mathcal{P}_{\mathbf{k},\mathbf{k}'}\left(\Phi_{\mathbf{k}}-\Phi_{\mathbf{k}'}\right),\label{eq:linColIntHere}\end{equation}
where for scattering by one in-plane phonon \begin{equation}
\mathcal{P}_{\mathbf{k},\mathbf{k}'}=\frac{2\pi}{\hbar}\sum_{\mathbf{q},\nu}w_{\nu}(\mathbf{q},\mathbf{k},\mathbf{k}')\omega_{\mathbf{q}}^{\nu}\frac{\partial n_{\mathbf{q}}}{\partial\omega_{\mathbf{q}}^{\nu}}\frac{\partial f_{\mathbf{k}}^{(0)}}{\partial\varepsilon_{\mathbf{k}}}\delta_{\mathbf{k},\mathbf{k}'+\mathbf{q}}\delta(\varepsilon_{\mathbf{k}}-\varepsilon_{\mathbf{k}'}),\label{eq:P1ph}\end{equation}
and for scattering by two FPs\begin{align}
\mathcal{P}_{\mathbf{k},\mathbf{k}'} & =-\frac{2\pi}{\hbar^{2}}k_{B}T\frac{\partial f_{\mathbf{k}}^{(0)}}{\partial\varepsilon_{\mathbf{k}}}\sum_{\mathbf{q},\mathbf{q}'}w_{F}(\mathbf{q},\mathbf{q}',\mathbf{k},\mathbf{k}')\frac{\partial n_{\mathbf{q}}}{\partial\omega_{\mathbf{q}}^{F}}\frac{\partial n_{\mathbf{q}'}}{\partial\omega_{\mathbf{q}'}^{F}}\times\nonumber \\
 & \times\left(\frac{\omega_{\mathbf{q}}^{F}+\omega_{\mathbf{q}'}^{F}}{1+n_{\mathbf{q}}+n_{\mathbf{q}'}}-\frac{\omega_{\mathbf{q}}^{F}-\omega_{\mathbf{q}'}^{F}}{n_{\mathbf{q}}-n_{\mathbf{q}'}}\right)\delta_{\mathbf{k},\mathbf{k}'+\mathbf{q}+\mathbf{q}'}\delta(\varepsilon_{\mathbf{k}}-\varepsilon_{\mathbf{k}'}),\label{eq:P2ph}\end{align}
with $n_{\mathbf{q}}=1/[\exp(\hbar\omega_{\mathbf{q}}/k_{B}T)-1]$
the equilibrium phonon distribution. The kernel quantities $w_{\nu}(\mathbf{q},\mathbf{k},\mathbf{k}')$
and $w_{F}(\mathbf{q},\mathbf{q}',\mathbf{k},\mathbf{k}')$ are related
to the matrix elements in Eq.~\eqref{eq:potential} as follows (see
Appendix~\ref{sec:APPcollInt}): for bilayer graphene,\begin{align}
w_{\nu}(\mathbf{q},\mathbf{k},\mathbf{k}') & \approx\left|V_{1,\mathbf{q}}^{\nu}\right|^{2}(1+\cos2\theta_{\mathbf{k},\mathbf{k}'})\nonumber \\
 & +\left|\tilde{V}_{2,\mathbf{q}}^{\nu}\right|^{2}\left(k^{2}+k'^{2}+2kk'\cos\theta_{\mathbf{k},\mathbf{k}'}\right),\label{eq:wKernelLT2L}\end{align}
for one phonon processes, and a similar expression for two phonon
processes $w_{F}(\mathbf{q},\mathbf{q}',\mathbf{k},\mathbf{k}')$
with $V_{\mathbf{q}}^{\nu}\rightarrow V_{\mathbf{q},\mathbf{q}'}^{F}$,
where $\tilde{V}_{2}$ means the matrix elements given in Eq.~\eqref{eq:potential}
for the gauge potential without the term $(\pi_{\mathbf{k}}+\pi_{\mathbf{k}'})$;
for monolayer graphene, in the case of one phonon process, \begin{equation}
w_{\nu}(\mathbf{q},\mathbf{k},\mathbf{k}')\approx\left|V_{1,\mathbf{q}}^{\nu}\right|^{2}(1+\cos\theta_{\mathbf{k},\mathbf{k}'})+\left|V_{2,\mathbf{q}}^{\nu}\right|^{2},\label{eq:wKernelLT1L}\end{equation}
with a similar expression for two phonon processes $w_{F}(\mathbf{q},\mathbf{q}',\mathbf{k},\mathbf{k}')$
with $V_{\mathbf{q}}^{\nu}\rightarrow V_{\mathbf{q},\mathbf{q}'}^{F}$.

Using the setting given above the resistivity is conveniently written
as\begin{align}
\varrho & =\frac{1}{g_{s}g_{v}}\frac{\frac{1}{2}\sum_{\mathbf{k},\mathbf{k}'}\left(\Phi_{\mathbf{k}}-\Phi_{\mathbf{k}'}\right)^{2}\mathcal{P}_{\mathbf{k},\mathbf{k}'}}{\mathcal{V}\left|\frac{e}{\mathcal{V}}\sum_{\mathbf{k}}\Phi_{\mathbf{k}}\mathbf{v}_{\mathbf{k}}\frac{\partial f_{\mathbf{k}}^{(0)}}{\partial\varepsilon_{\mathbf{k}}}\right|^{2}}\nonumber \\
 & \approx\frac{\mathcal{V}}{8e^{2}}\frac{\int d\mathbf{k}d\mathbf{k}'\left(\mathbf{K}\cdot\mathbf{u}\right)^{2}\mathcal{P}_{\mathbf{k},\mathbf{k}'}}{\left|\int d\mathbf{k}\,\mathbf{k}\cdot\mathbf{u}\mathbf{v}_{\mathbf{k}}\frac{\partial f_{\mathbf{k}}^{(0)}}{\partial\varepsilon_{\mathbf{k}}}\right|^{2}},\label{eq:resVarGraphene}\end{align}
where we changed from summation over $\mathbf{k}-$space to integration,
and defined $\mathbf{K}=\mathbf{k}-\mathbf{k}'$. The integral in
the denominator can be done immediately assuming $\varepsilon_{F}\gg k_{B}T$.
The result reads the same for bilayer and monolayer graphene, \begin{equation}
\left|\int d\mathbf{k}\,\mathbf{k}\cdot\mathbf{u}\mathbf{v}_{\mathbf{k}}\frac{\partial f_{\mathbf{k}}^{(0)}}{\partial\varepsilon_{\mathbf{k}}}\right|\approx\frac{\pi k_{F}^{2}}{\hbar}.\label{eq:current}\end{equation}
In order to proceed analytically with the integral in the numerator
we have to specify the $T$ regime, as discussed in the next section.

\subsection{Bloch -Gr\"uneisen temperature}

For each scattering process (one or two phonon scattering) we may
identify two different $T$ regimes, \emph{low} and \emph{high} $T$,
related to whether only small angle or every angle are available to
scatter from $\bigl|\mathbf{k}\bigr\rangle$ to $\bigl|\mathbf{k}'\bigr\rangle$.
Recall that since we are dealing with quasielastic scattering both
$\mathbf{k}$ and \textbf{$\mathbf{k}'$} sit on the Fermi circle,
see Fig.~\ref{fig:scattDraw}, and $\bigl|\mathbf{k}\bigr\rangle$
and $\bigl|\mathbf{k}'\bigr\rangle$ are adiabatically connected through
a rotation of $\theta_{\mathbf{k},\mathbf{k}'}$ in momentum space.
Large angle scattering is only possible if phonons with high enough
momentum are available to scatter electrons. The characteristic Bloch-Gr\"uneisen
temperature $T_{BG}$ separating the two regimes may thus be set by
the minimum phonon energy necessary to have full back scattering,\begin{equation}
k_{B}T_{BG}=\hbar\omega_{2k_{F}},\label{eq:TBG}\end{equation}
with $\omega_{\mathbf{q}}$ as given in Sec.~\ref{sec:phonons}.

For scattering by in-plane phonons $T_{BG}$ takes the value \begin{equation}
T_{BG}^{(L)}\approx57\sqrt{n}\,\mbox{K}\,\,\,\,\,\,\mbox{and}\,\,\,\,\,\, T_{BG}^{(T)}\approx38\sqrt{n}\,\mbox{K},\label{eq:TBGin}\end{equation}
 for longitudinal and transverse phonons respectively, with density
$n$ in units of $10^{12}\,\mbox{cm}^{-2}$. When scattering is by
two non-strained FPs, the crossover between low and high $T$ regimes
is given by \begin{equation}
T_{BG}\approx0.1n\,\mbox{K},\label{eq:TBGunStrF}\end{equation}
with $n$ again measured in $10^{12}\,\mbox{cm}^{-2}$, while in the
presence of strain, using the approximated strained FP dispersion
in Eq.~\eqref{eq:flexDispStr4}, we get \begin{equation}
T_{BG}\simeq28\sqrt{n\bar{u}}\,\,\mbox{K}.\label{eq:TBGstrF}\end{equation}
It is obvious from Eqs.~\eqref{eq:TBGunStrF} and~\eqref{eq:TBGstrF}
that the high-$T$ regime is the relevant one for FP scattering.

\subsection{Contributions to resistivity}

\label{sub:ResContrib}

In the following we summarize our results for the $T$ dependent resistivity
due to scattering by in-plane phonons and two FPs in bilayer graphene.
For comparison we discuss also the monolayer case first studied in
Ref.~\onlinecite{MvO10}. We use the variational method discussed
in Sec.~\ref{sub:varMethod}; the resistivity being given by Eq.~\eqref{eq:resVarGraphene}.
Details on the derivation can be found in Appendix~\ref{sec:APPres}.
We neglect one FP processes since these, as can be seen in Eq.~\eqref{eq:potential},
are reduced by a factor $\sim h_{0}/L\ll1$, where $h_{0}$ is the
sample's vertical deflection over the typical linear size $L$.

\subsubsection{Scattering by in-plane phonons}

\label{sub:ResContribInP}

A sketch of the scattering process in momentum space involving one
phonon is shown in Fig.~\ref{fig:scattDraw}(a). In this case the
resistivity can be written as (see Appendix~\ref{sub:APPresInPlane}),\begin{equation}
\varrho_{in}\approx\frac{8\hbar k_{F}^{2}}{e^{2}\rho v_{F}^{2}k_{B}T}\sum_{\nu}\int_{0}^{1}dx\,[D_{B}^{\nu}(2x)]^{2}\frac{x^{4}}{\sqrt{1-x^{2}}}\frac{e^{xz_{\nu}}}{(e^{xz_{\nu}}-1)^{2}},\label{eq:resIn}\end{equation}
where $z_{\nu}=\hbar\omega_{2k_{F}}^{\nu}/k_{B}T$, with $\nu=L,T$,
and where we have introduced a generalized electron--in-plane phonon
coupling for bilayer graphene given by,

\begin{equation}
D_{B}^{\nu}(y)=\left[2g^{2}y^{2}\left(1-\frac{y^{2}}{2}\right)^{2}\delta_{\nu L}+\frac{\hbar^{2}v_{F}^{2}\beta^{2}}{4a^{2}}\left(1-\frac{y^{2}}{4}\right)\right]^{1/2}.\label{eq:couplingBilIn}\end{equation}
The case of scattering via screened scalar potential, here encoded
in the screened deformation potential parameter $g$, has been considered
recently in Ref.~\onlinecite{MHdS10}. As it is shown below, the gauge
potential contribution becomes the dominant one in the low $T$ regime.

\begin{figure}
\begin{centering}
\includegraphics[width=0.7\columnwidth]{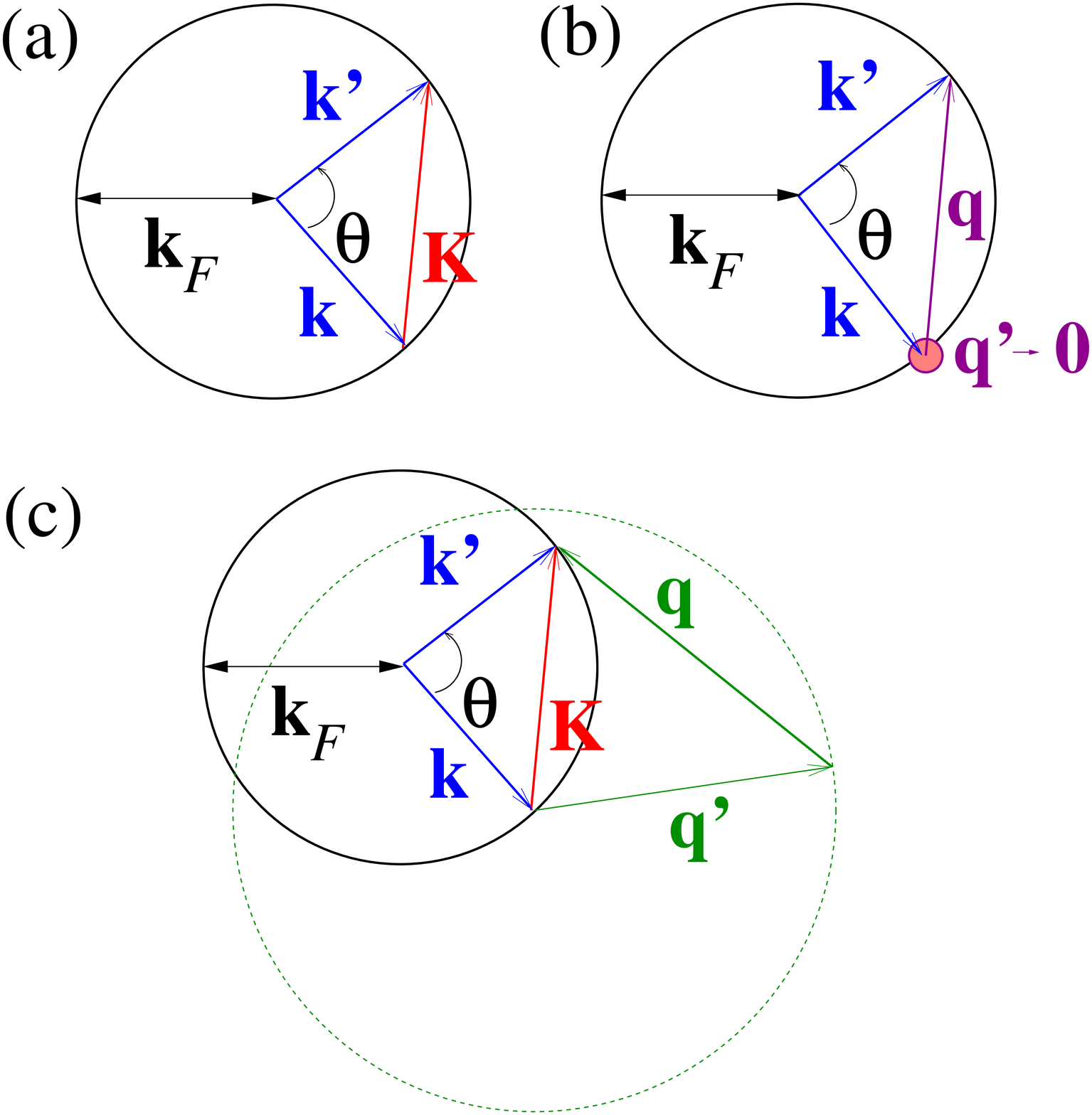}
\par\end{centering}

\caption{\label{fig:scattDraw}Scattering of electrons in momentum space due
to: (a)~in-plane phonons, (b)~non-strained flexural and (c)~strained
flexural phonons.}

\end{figure}

In the low $T$ regime, $T\ll T_{BG}$, we have $z_{\nu}\gg1$, so
that the integrand in Eq.~\eqref{eq:resIn} is only contributing
significantly for $x\ll1$. The generalized electron--in-plane phonon
coupling in~\eqref{eq:couplingBilIn} then becomes,\begin{equation}
D_{B}^{\nu}(y\ll1)=\left[2g^{2}y^{2}\delta_{\nu L}+\frac{\hbar^{2}v_{F}^{2}\beta^{2}}{4a^{2}}\right]^{1/2},\label{eq:couplingLowTIn}\end{equation}
and the resistivity reads,\begin{multline}
\varrho_{in}\approx\sum_{\nu}\left[g^{2}\frac{16\Gamma(6)\zeta(6)}{\Gamma(4)\zeta(4)}\left(\frac{T}{T_{BG}}\right)^{2}\delta_{\nu L}+\frac{\hbar^{2}v_{F}^{2}\beta^{2}}{4a^{2}}\right]\times\\
\frac{\Gamma(4)\zeta(4)(k_{B}T)^{4}}{e^{2}\rho\hbar^{4}v_{F}^{2}v_{\nu}^{5}k_{F}^{3}},\label{eq:resLowTIn}\end{multline}
where $\Gamma(n)=(n-1)!$ is the \emph{gamma function} and $\zeta(n)$
is the \emph{Riemann zeta function}. We have thus obtained the expected
$T^{4}$ behavior at low $T$ for coupling through gauge potential,
which is the 2--dimensional analogue of the $T^{5}$ Bloch theory
in 3--dimensional metals.\cite{zimanEP,HdSacph07} The scalar potential
contribution comes proportional to $T^{6}$ due to screening. It can
be neglected in the low $T$ regime; even though $16\Gamma(6)\zeta(6)/[\Gamma(4)\zeta(4)]\approx300$,
it is strongly suppressed by $T/T_{BG}\ll1$ and $g<\hbar v_{F}\beta/(2a)$,
(see Sec.~\ref{sub:coupling}).

In the high $T$ regime, $T\gg T_{BG}$, the inequality $z_{\nu}\ll1$
holds, so that $e^{z_{\nu}x}/(e^{z_{\nu}x}-1)^{2}\approx1/(z_{\nu}x)$
in Eq.~\eqref{eq:resIn}. The usual linear in $T$ resistivity for
one phonon scattering is then recovered,\begin{equation}
\varrho_{in}\approx\left(7g^{2}+\frac{\hbar^{2}v_{F}^{2}\beta^{2}}{8a^{2}}\frac{v_{L}^{2}}{\bar{v}^{2}}\right)\frac{\pi k_{B}T}{4\hbar\rho e^{2}v_{L}^{2}v_{F}^{2}},\label{eq:resHighTIn}\end{equation}
where $1/\bar{v}^{2}=1/v_{L}^{2}+1/v_{T}^{2}$. Note that, at odds
with the low $T$ regime, now the scalar potential contribution is
higher than the gauge potential one for the typical coupling values
discussed in Sec.~\ref{sub:coupling}.

The monolayer case has been discussed extensively in the literature.\cite{HdSacph07,SPG07,MO08,MvO10,COK+10}
The resistivity is still given by Eq.~\eqref{eq:resIn}, only the
generalized electron--in-plane phonon coupling changes,\begin{equation}
D_{M}^{\nu}(y)=\left[2g^{2}y^{2}\left(1-\frac{y^{2}}{4}\right)\delta_{\nu L}+\frac{\hbar^{2}v_{F}^{2}\beta^{2}}{4a^{2}}\right]^{1/2}.\label{eq:couplingMonoIn}\end{equation}
The same qualitative behavior is obtained: at low $T$ the resistivity
is given by Eq.~\eqref{eq:resLowTIn} with the numerical replacement
$16\rightarrow12$ in the scalar potential contribution; at high $T$
the result~\eqref{eq:resHighTIn} holds with the replacement $\left(7g^{2}+\frac{\hbar^{2}v_{F}^{2}\beta^{2}}{8a^{2}}\frac{v_{L}^{2}}{\bar{v}^{2}}\right)\rightarrow\left(2g^{2}+\frac{\hbar^{2}v_{F}^{2}\beta^{2}}{2a^{2}}\frac{v_{L}^{2}}{\bar{v}^{2}}\right)$.
Note, however, the apparent quantitative difference: the scalar and
gauge potential contributions change roles, the later becoming more
important in monolayer graphene. This is further discussed in Sec.~\ref{sec:Discuss}.

A final remark regarding the temperature dependent resistivity due
to in-plane phonons has to do with the value of the electron-phonon
coupling parameters $\beta$ and $g$. While $\beta$ is expected
to be restricted to the range $\beta\sim2-3$, as discussed in Sec.~\ref{sub:coupling},
the value of the deformation potential parameter $g$ is still debated
in the literature. Phenomenology gives $g\sim10-30\,\,\mbox{eV}$;\cite{SA02a,HdSacph07}
recent \emph{ab initio} calculations provide a much smaller value
$g\sim3\,\,\mbox{eV}$.\cite{CJS10} On the other hand, experiments
seem to confirm the higher values, giving $g\sim15-25\,\,\mbox{eV}$.\cite{CXI+07,EK10}
Our claim here is that all these values make sense, if properly interpreted:
phenomenology gives essentially unscreened deformation potential,
which we called $g_{0}$ in Sec.~\ref{sub:coupling}, and which should
take values of $\mathcal{O}(10)\,\,\mbox{eV}$; screening effects
suppress the deformation potential to $\mathcal{O}(1)\,\,\mbox{eV}$,
as we have seen in Sec.~\ref{sub:coupling} within the Thomas Fermi
approximation, in good agreement with \emph{ab initio} results where
screening is built in; the fact that transport experiments give a
much higher deformation potential is a strong indication that phonon
scattering through gauge potential, usually not included when fitting
the data,\cite{CXI+07,EK10} is at work. Indeed, using the monolayer
version of Eq.~\eqref{eq:resHighTIn}, we readily find that the fitting
quantity in Refs.~\onlinecite{CXI+07} and~\onlinecite{EK10} should
be replaced by,\begin{equation}
\tilde{D}=\left[2g^{2}+\frac{v_{F}^{2}\hbar^{2}\beta^{2}}{2a^{2}}\left(1+\frac{v_{L}^{2}}{v_{T}^{2}}\right)\right]^{1/2},\label{eq:goodFit}\end{equation}
which, keeping $g\sim3\,\,\mbox{eV}$, takes values $\tilde{D}\sim10-20\,\,\mbox{eV}$
for $\beta\sim2-3$, in excellent agreement with experiments. Moreover,
since the gauge potential is not screened {[}Eq.~\eqref{eq:resLowTIn}{]}
it provides a natural explanation for the $T^{4}$ resistivity behavior
recently reported at low $T$ in Ref.~\onlinecite{EK10}, where the
expected $T^{6}$ contribution due to scalar potential is absent.\cite{MHdS10}

Typical $T$ dependent resistivity due to scattering by in-plane phonons,
Eq.~\eqref{eq:resIn}, is shown in Fig.~\ref{fig:res} as thick
dashed lines. In agreement with analytical results, there is no qualitative
difference between monolayer graphene {[}Fig.~\ref{fig:res}(a){]}
and bilayer {[}Fig.~\ref{fig:res}(b){]}.

\subsubsection{Scattering by non-strained flexural phonons}

\label{sub:resNstrFP}

\begin{figure}
\begin{centering}
\includegraphics[width=0.95\columnwidth]{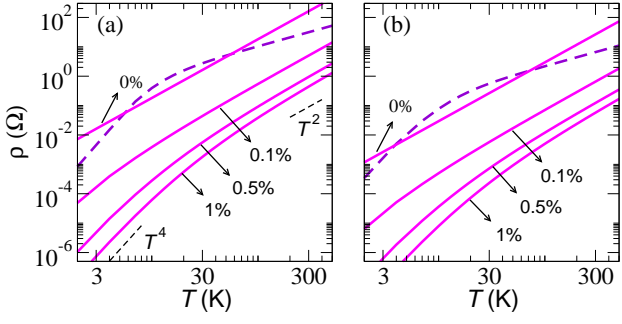}
\par\end{centering}

\caption{\label{fig:res}Resistivity vs $T$ due to scattering by in-plane
phonons (dashed thick lines) and flexural phonons (full lines) with
and without strain (indicated in percentage), in monolayer~(a) and
bilayer graphene~(b). We use $n=10^{12}\,\mbox{cm}^{-2}$, $g\approx3\,\mbox{eV}$,
and $\beta\approx3$.}

\end{figure}

In non-strained bilayer graphene scattering by FPs give rise to the
following $T$ dependent resistivity (details on the derivation are
given in Appendix~\ref{sub:APPresFlex}),\begin{equation}
\varrho_{F}\approx\frac{\hbar k_{F}^{2}}{2\pi e^{2}\rho^{2}v_{F}^{2}\alpha^{2}}\ln\left(\frac{k_{B}T}{\hbar\alpha q_{c}^{2}}\right)\int_{0}^{1}dx\frac{[D_{B}^{F}(2x)]^{2}}{\sqrt{1-x^{2}}}\frac{x^{4}e^{zx^{2}}}{(e^{zx^{2}}-1)^{2}},\label{eq:resNsFlex}\end{equation}
where $z=\hbar\omega_{2k_{F}}^{F}/k_{B}T$, and where the generalized
electron--FP coupling for bilayer graphene is given by,\begin{equation}
D_{B}^{F}(x)=\left[g^{2}x^{2}\left(1-\frac{x^{2}}{2}\right)^{2}+\frac{\hbar^{2}v_{F}^{2}\beta^{2}}{4a^{2}}\left(1-\frac{x^{2}}{4}\right)\right]^{1/2}.\label{eq:couplingBilFlex}\end{equation}
Equation~\eqref{eq:resNsFlex} holds also for monolayer graphene,
we need only to introduce a different generalized electron-FP coupling,\begin{equation}
D_{M}(x)=\left[g^{2}x^{2}\left(1-\frac{x^{2}}{4}\right)+\frac{\hbar^{2}v_{F}^{2}\beta^{2}}{4a^{2}}\right]^{1/2}.\label{eq:couplingMonFlex}\end{equation}
In Fig.~\ref{fig:scattDraw}(b) a sketch of the two phonon scattering
process in momentum space is provided. It shows that one of the two
phonons involved in the scattering event always has momentum $\mathbf{q}'\rightarrow0$.
This is a consequence of the quadratic FP dispersion {[}Eq.~\eqref{eq:flexDispUnStr}{]},
which leads to a divergent number of FPs with momentum $\mathbf{q}'\rightarrow0$.\cite{NGPrmp}
This divergence is responsible for the logarithmic factor in Eq.~\eqref{eq:resNsFlex},
which stems from the existence of an infrared cutoff $q_{c}$. This
cutoff is to be identified with the onset of anharmonic effects,\cite{ZRF+10}
or unavoidable built in strain.\cite{COK+10}

In the low $T$ regime, $T\ll T_{BG}$, one has $z\gg1$, so that
the integrand in Eq.~\eqref{eq:resNsFlex} is only contributing for
$x\ll1$. The generalized electron-phonon coupling becomes equal in
both bilayer and monolayer systems,\begin{equation}
D_{B}(y\ll1)=D_{M}(y\ll1)=\left[g^{2}y^{2}+\frac{\hbar^{2}v_{F}^{2}\beta^{2}}{4a^{2}}\right]^{1/2},\label{eq:couplingLowT}\end{equation}
and the resistivity is then the same in both,\begin{multline}
\varrho_{F}\approx\left[g^{2}\frac{6\Gamma(6)\zeta(6)}{\Gamma(4)\zeta(4)}\left(\frac{T}{T_{BG}}\right)^{2}+\frac{\hbar^{2}v_{F}^{2}\beta^{2}}{4a^{2}}\right]\times\\
\frac{\Gamma(4)\zeta(4)\hbar k_{F}^{2}}{2^{4}\pi e^{2}\rho^{2}v_{F}^{2}\alpha^{2}}\left(\frac{k_{B}T}{\hbar\alpha k_{F}^{2}}\right)^{5/2}\ln\left(\frac{k_{B}T}{\hbar\alpha q_{c}^{2}}\right).\label{eq:resNsFlexLowT}\end{multline}
A similar result has been derived in Ref.~\onlinecite{MO08}. Owing
to the same arguments used in the previous section for one phonon
scattering we can neglect the scalar potential contribution at low
$T$.

At high $T$, i.e. $T\gg T_{BG}$, we have $z\ll1$, so that $\exp(zx^{2})/[\exp(zx^{2})-1]^{2}\approx1/(zx^{2})$
in Eq.~\eqref{eq:resNsFlex}. The bilayer graphene resistivity becomes,\begin{equation}
\varrho_{F}\approx\left(g^{2}+\frac{\hbar^{2}v_{F}^{2}\beta^{2}}{8a^{2}}\right)\frac{(k_{B}T)^{2}}{64\hbar e^{2}\rho^{2}v_{F}^{2}\alpha^{4}k_{F}^{2}}\ln\left(\frac{k_{B}T}{\hbar\alpha q_{c}^{2}}\right).\label{eq:resNsFlexHighT}\end{equation}
This result holds for monolayer graphene with the substitution $\left(g^{2}+\frac{\hbar^{2}v_{F}^{2}\beta^{2}}{8a^{2}}\right)\rightarrow\left(\frac{g^{2}}{2}+\frac{\hbar^{2}v_{F}^{2}\beta^{2}}{4a^{2}}\right)$.\cite{MvO10,COK+10}
We have obtained that the resistivity due to non-strained FPs is proportional
to $T^{2}/n$, which implies mobility independent of the carrier density
$n$. A similar result has been obtained in the context of microscopic
ripples in graphene.\cite{KG08,MNK+07} The result of Eq.~\eqref{eq:resNsFlexHighT}
is shown in Fig.~\ref{fig:res}(b) as a full line indicating $\bar{u}\approx0\%$.
The logarithmic correction, expected to be of order unity in the relevant
$T$ range, has been ignored. Scattering by FPs dominates the contribution
to resistivity in non-strained samples at both low and high $T$,
except for the crossover region where $T\sim T_{BG}$, Eq.~\eqref{eq:TBGin}.
The same conclusion holds for monolayer graphene, whose $T$ dependent
$\varrho_{F}$ is shown in Fig.~\ref{fig:res}(a).

\subsubsection{Scattering by strained flexural phonons}

Applying strain breaks the membrane rotational symmetry inducing linear
FP dispersion at low momentum, as can be seen in Eq.~\eqref{eq:flexDispStr4}.
A new energy scale appears in the problem, \begin{equation}
\omega_{q^{*}}^{F}=\sqrt{2}\bar{u}v_{L}^{2}/\alpha\approx10^{4}\bar{u}(\mbox{K}),\label{eq:newEnScale}\end{equation}
separating two regimes: linear dispersion below and quadratic dispersion
above. The associated momentum scale, $q^{*}=\sqrt{\bar{u}}v_{L}/\alpha\approx4.5\sqrt{\bar{u}}\,\,\mbox{\AA}^{-1}$,
together with $k_{F}$ and the thermal momentum $q_{T}$ given by
$\hbar\omega_{q_{T}}^{F}=k_{B}T$, define all regimes where analytic
treatment can be employed. In particular, in the low $T$ regime where
$q_{T}\ll k_{F}$ we may always take $q^{*}\gg q_{T}$ and use a linear
dispersion for FPs; otherwise the non strained case considered in
the previous section would be the appropriate starting point. In the
high $T$ regime we can distinguish between low strain for $q^{*}\ll q_{T}$
and high strain for $q^{*}\gg q_{T}$. Note that at high $T$ relevant
phonons scattering electrons have momentum $q$ in the range $k_{F}\lesssim q\lesssim q_{T}$.
Therefore, when strain is present in the high $T$ regime we may always
assume $q^{*}\gg k_{F}$; the opposite limit, $q^{*}\ll k_{F}$, would
again be identified with the non-strained case considered previously.

The resistivity due to strained FPs can be cast in the form of a triple
integral over rescaled momenta $x\rightarrow\tilde{x}=x\hbar v_{L}u^{1/2}/(k_{B}T)$
(see Appendix~\ref{sub:APPresFlex} for details),\begin{widetext}\begin{multline}
\varrho_{F}\approx\frac{(k_{B}T)^{6}}{2^{6}\pi^{2}\hbar^{5}e^{2}\rho^{2}v_{F}^{2}v_{L}^{8}\bar{u}^{4}k_{F}^{2}}\int_{0}^{2\tilde{k}_{F}}d\tilde{K}\frac{[D_{B}^{F}(\tilde{K}/\tilde{k}_{F})]^{2}\tilde{K}^{2}}{\sqrt{\tilde{k}_{F}^{2}-\tilde{K}^{2}/4}}\int_{0}^{\infty}d\tilde{q}\,\frac{\tilde{q}^{3}}{\omega_{\tilde{q}}}n_{\tilde{q}}(n_{\tilde{q}}+1)\times\\
\int_{|\tilde{K}-\tilde{q}|}^{|\tilde{K}+\tilde{q}|}d\tilde{Q}\frac{\tilde{Q}^{3}n_{\tilde{Q}}(n_{\tilde{Q}}+1)}{\omega_{\tilde{Q}}\sqrt{\tilde{q}^{2}\tilde{K}^{2}-\Bigl(\tilde{K}^{2}+\tilde{q}^{2}-\tilde{Q}^{2}\Bigr)^{2}/4}}\left(\frac{\omega_{\tilde{q}}+\omega_{\tilde{Q}}}{1+n_{\tilde{q}}+n_{\tilde{Q}}}-\frac{\omega_{\tilde{q}}-\omega_{\tilde{Q}}}{n_{\tilde{q}}-n_{\tilde{Q}}}\right),\label{eq:resStrFlex}\end{multline}
\end{widetext}where the rescaled dispersion reads $\omega_{\tilde{x}}\approx\sqrt{\gamma^{2}\tilde{x}^{4}+\tilde{x}^{2}}$,
with $\gamma=\sqrt{2}\omega_{q_{T}}^{F}/\omega_{q^{*}}^{F}$, and
the generalized electron-FP coupling $D_{B}^{F}(y)$ is given by Eq.~\eqref{eq:couplingBilFlex}.
For monolayer graphene only the coupling changes, being given instead
by Eq.~\eqref{eq:couplingMonFlex}. The kinematics of the scattering
process is schematically shown in Fig.~\ref{fig:scattDraw}(c).

In the low $T$ case, $T\ll T_{BG}$, we have only small angle scattering
with $K\ll k_{F}$. The argument of the generalized electron-FP coupling
becomes small, $K/k_{F}\ll1$, and it can be written as\begin{equation}
D_{B}(y\ll1)\approx D_{M}(y\ll1)\approx\left[g^{2}y^{2}+\frac{\hbar^{2}v_{F}^{2}\beta^{2}}{4a^{2}}\right]^{1/2}.\label{eq:couplingFlexLowT}\end{equation}
The resistivity is the same in both bilayer and monolayer systems.
Since the inequality $q_{T}\ll k_{F},q^{*}$ holds, relevant phonons
have linear dispersion $\omega_{q}^{F}\approx\sqrt{\bar{u}}v_{L}q$
and the rescaled Fermi momentum obeys $\tilde{k}_{F}\approx k_{F}/q_{T}\gg1$.
We may take $\tilde{K}\rightarrow\infty$ as the upper limit in the
$\tilde{K}$ integral in Eq.~\eqref{eq:resStrFlex}, and the resistivity
is then approximated by,\begin{equation}
\varrho_{F}\approx\left[g^{2}\left(\frac{q_{T}}{k_{F}}\right)^{2}\mathcal{K}_{4}+\frac{\hbar^{2}v_{F}^{2}\beta^{2}}{4a^{2}}\mathcal{K}_{2}\right]\frac{(k_{B}T)^{7}}{2^{6}\pi^{2}\hbar^{6}e^{2}\rho^{2}v_{F}^{2}v_{L}^{9}\bar{u}^{9/2}k_{F}^{3}},\label{eq:resFlexStrLowT}\end{equation}
where\begin{multline}
\mathcal{K}_{n}=\int_{0}^{\infty}d\tilde{K}\tilde{K}^{n}\int_{0}^{\infty}d\tilde{q}\, q^{2}n_{\tilde{q}}(n_{\tilde{q}}+1)\times\\
\int_{|\tilde{K}-\tilde{q}|}^{|\tilde{K}+\tilde{q}|}d\tilde{Q}\frac{\tilde{Q}^{2}n_{\tilde{Q}}(n_{\tilde{Q}}+1)\left(\frac{\tilde{q}+\tilde{Q}}{1+n_{\tilde{q}}+n_{\tilde{Q}}}-\frac{\tilde{q}-\tilde{Q}}{n_{\tilde{q}}-n_{\tilde{Q}}}\right)}{\sqrt{\tilde{q}^{2}\tilde{K}^{2}-\Bigl(\tilde{K}^{2}+\tilde{q}^{2}-\tilde{Q}^{2}\Bigr)^{2}/4}}.\label{eq:lowTkernel}\end{multline}
It can be shown numerically that $\mathcal{K}_{2}\approx4485$ and
$\mathcal{K}_{4}\approx496850$. The large ratio $\mathcal{K}_{4}/\mathcal{K}_{2}\gg1$
is, however, compensated by $q_{T}/k_{F}\ll1$ and the fact that $g<\hbar v_{F}\beta/(2a)$
(see Sec.~\ref{sub:coupling}). As in the case of scattering by in-plane
phonons, also here the gauge potential contribution to resistivity
dominates at low $T$.

Now we consider the high $T$ regime, $T\gg T_{BG}$. At odds with
the non-strained case {[}see Fig.~\ref{fig:scattDraw}(b){]}, now
phonons with momentum $q$ in the range $k_{F}\lesssim q\lesssim q_{T}$
provide most of the scattering. It is shown in Appendix~\ref{sub:APPresFlexS}
that the integral over $\tilde{Q}$ in Eq.~\eqref{eq:resStrFlex}
becomes $\tilde{K}$ independent, and the $\tilde{q}$ integral can
be cast in the form,\begin{equation}
\mathcal{G}(\gamma)=\int_{0}^{\infty}d\tilde{q}\,\frac{\tilde{q}^{5}n_{\tilde{q}}^{2}(n_{\tilde{q}}+1)^{2}}{\gamma^{2}\tilde{q}^{4}+\tilde{q}^{2}}\left(\frac{2\sqrt{\gamma^{2}\tilde{q}^{4}+\tilde{q}^{2}}}{1+2n_{\tilde{q}}}+\frac{1}{n_{\tilde{q}}(n_{\tilde{q}}+1)}\right),\label{eq:Ggamma}\end{equation}
being easily evaluated numerically. The resistivity in bilayer graphene
can then be written as\begin{equation}
\varrho_{F}\approx\left(7g^{2}+\frac{\hbar^{2}v_{F}^{2}\beta^{2}}{4a^{2}}\right)\frac{(k_{B}T)^{4}}{2^{7}\hbar^{3}e^{2}\rho^{2}v_{F}^{2}v_{L}^{6}\bar{u}^{3}}\mathcal{G}\left(\frac{\alpha k_{B}T}{\hbar\bar{u}v_{L}^{2}}\right).\label{eq:resFlexStrHighT}\end{equation}
When $\gamma\ll1$, or equivalently $q_{T}\ll q^{*}$ (high strain),
the function $\mathcal{G}(\gamma)$ behaves as $\mathcal{G}(\gamma\ll1)\approx18\zeta(3)-93\zeta(5)/8$.
For $\gamma\gg1$, or equivalently $q_{T}\gg q^{*}$ (small strain),
it gives $\mathcal{G}(\gamma\gg1)\approx1/\gamma^{2}$. In these asymptotic
regimes one can obtain analytic expressions for the resistivity in
Eq.~\eqref{eq:resFlexStrHighT},\begin{multline}
\varrho_{F}\approx\left(7g^{2}+\frac{\hbar^{2}v_{F}^{2}\beta^{2}}{4a^{2}}\right)\frac{1}{2^{7}\hbar^{3}e^{2}v_{F}^{2}}\times\\
\begin{cases}
\left(18\zeta(3)-\frac{93}{8}\zeta(5)\right)\frac{(k_{B}T)^{4}}{\rho^{2}v_{L}^{6}\bar{u}^{3}} & k_{F}\ll q_{T}\ll q^{*}\\
\frac{\hbar^{2}(k_{B}T)^{2}}{\rho\kappa v_{L}^{2}\bar{u}} & k_{F}\ll q^{*}\ll q_{T}\end{cases}.\label{eq:resFlexStrHighTanal}\end{multline}
Equations~\eqref{eq:resFlexStrHighT} and~\eqref{eq:resFlexStrHighTanal}
also hold for monolayer graphene with $\left(7g^{2}+\frac{\hbar^{2}v_{F}^{2}\beta^{2}}{4a^{2}}\right)\rightarrow\left(2g^{2}+\frac{\hbar^{2}v_{F}^{2}\beta^{2}}{a^{2}}\right)$.

The effect of strain in the $T$ dependence of resistivity is shown
in Fig.~\ref{fig:res}(a) for monolayer graphene and~\ref{fig:res}(b)
for bilayer at strain values $\bar{u}\approx0.1\%,0.5\%,1\%$. The
crossover between the two regimes of Eq.~\eqref{eq:resFlexStrHighT}
{[}see Eq.~\eqref{eq:resFlexStrHighTanal}{]} is clearly seen at
$\gamma\approx1$, which is equivalent to $T\approx10^{4}\bar{u}\,\,\mbox{K}$.
It is apparent from Fig.~\ref{fig:res} that the contribution to
the resistivity due to scattering by FPs is strongly suppressed by
applying strain.

\subsection{Crossover between in-plane and flexural phonon dominated scattering}

Scattering by in-plane and flexural phonons are always at work simultaneously.
However, the two mechanisms provide completely different $T$ dependent
resistivity, and therefore we expect them to dominate at different
$T$. In the following we address the transition $T$ at which $\varrho_{in}\approx\varrho_{F}$.

\subsubsection{Non-strained case}

In this case, using Eqs.~\eqref{eq:resHighTIn} and~\eqref{eq:resNsFlexHighT}
for bilayer graphene in the high $T$ regime we get\begin{equation}
\frac{\varrho_{F}}{\varrho_{in}}\approx\frac{T[K]}{75n[\mbox{cm}^{-2}]},\label{eq:resRatNoS2}\end{equation}
We expect a crossover between in-plane to FP dominated scattering
given by\begin{equation}
T_{c2}\approx75n[\mbox{cm}^{-2}]\,\,\mbox{K}.\label{eq:TcNoS2}\end{equation}
The $T_{c}$ just obtained is close to $T_{BG}$ for in-plane phonons,
Eq.~\eqref{eq:TBGin}, and much higher than $T_{BG}$ for FPs, Eq.~\eqref{eq:TBGunStrF}.
Using the low $T$ approximation for $\varrho_{in}$, Eq.~\eqref{eq:resLowTIn},
we obtain the ratio\begin{equation}
\frac{\varrho_{F}}{\varrho_{in}}\approx12\frac{\sqrt{n[\mbox{cm}^{-2}]}}{(T[\mbox{K}])^{2}},\label{eq:resRatNoS1}\end{equation}
from which we expect a crossover from FP to in-plane dominated scattering
at\begin{equation}
T_{c1}\approx3(n[\mbox{cm}^{-2}])^{1/4}\,\,\mbox{K},\label{eq:TcNoS1}\end{equation}
as $T$ increases. We conclude that scattering by FP always dominates
over scattering by in-plane ones, except for the region $T_{c1}\ll T\ll T_{c2}$
around $T_{BG}$ for in-plane phonons, Eq.~\eqref{eq:TBGin}. This
is clearly seen in Fig.~\ref{fig:res}(b). The same conclusion applies
to monolayer graphene. In this later case we obtain $T_{c1}\approx6(n[\mbox{cm}^{-2}])^{1/4}\,\,\mbox{K}$
and $T_{c2}\approx55n[\mbox{cm}^{-2}]\,\,\mbox{K}$.

\subsubsection{Strained case}

It can easily be shown that the crossover from in-plane to flexural
phonon dominated scattering always occurs in the low strain regime,
$q^{*}\ll q_{T}$. We have seen in the previous sections that the
crossover temperature $T$ separating high strain from low strain
behavior is given by $\gamma=\sqrt{2}\omega_{q_{T}}^{F}/\omega_{q^{*}}^{F}\approx1$.
Using Eq.~\eqref{eq:newEnScale} we get a crossover temperature $T^{*}\approx10^{4}\bar{u}\,\mbox{K}$.
On the other hand, using the low strain approximation for the resistivity
due to flexural phonons given in Eq.~\eqref{eq:resFlexStrHighTanal}
and the resistivity due to in-plane ones in Eq.~\eqref{eq:resHighTIn}
we obtain for the ratio in bilayer graphene \begin{equation}
\frac{\varrho_{F}}{\varrho_{in}}\approx\frac{k_{B}T}{40\pi\kappa\bar{u}}.\label{eq:resRatS}\end{equation}
The corresponding crossover $T$ then reads\begin{equation}
T_{c}\approx10^{6}\bar{u}\,\mbox{K}.\label{eq:TcS}\end{equation}
Clearly $T_{c}\gg T^{*}$, justifying our low strain approximation.
The same applies to monolayer graphene under strain. The resistivity
ratio is in that case $\varrho_{F}/\varrho_{in}\approx k_{B}T/(50\pi\kappa\bar{u})$,
from which we obtain roughly the same $T_{c}$ even taken into account
that $\kappa$ in monolayer is half that of bilayer in our elasticity
model.

In important conclusion may be drawn. While in the non-strained case
scattering by FP is the dominant contribution to the resistivity,
it can be seen from Eq.~\eqref{eq:TcS} that applying small amounts
of strain is enough to suppress this contribution at room $T$. This
is clearly seen in Fig.~\ref{fig:res}.

\section{Discussion}

\label{sec:Discuss}

We have found the $T$ dependent resistivity due to acoustic phonons
to be qualitatively similar in monolayer and bilayer graphene (see
Sec.~\ref{sub:ResContrib}). This becomes apparent when we compare
Fig.~\ref{fig:res}(a) and~\ref{fig:res}(b) where $\varrho(T)$
is shown, respectively, for monolayer and bilayer graphene both at
zero and finite strain. Such behavior can be traced back to the resistivity
expression, Eq.~\eqref{eq:resVarGraphene}, or more precisely to
its numerator, where the different electronic structure and electron-phonon
coupling conspire to give exactly the same parametric dependence in
monolayer and bilayer graphene. In short, it can be readily seen through
Eq.~\eqref{eq:resVarGraphene} that the information about the electronic
structure enters via the density of states squared in the numerator.
The electron-phonon coupling, in its turn, enters through the transition
rate $\mathcal{P}_{\mathbf{k},\mathbf{k}'}$, and in the Born approximation
it also appears squared. When coupling is via scalar potential {[}$V_{1}$
matrix elements in Eq.~\eqref{eq:potential}{]} the screening makes
the coupling inversely proportional to the density of states, and
the square of it cancels exactly with the square density of states
coming from the integral over $\mathbf{k}$ and $\mathbf{k}'$ in
the numerator of Eq.~\eqref{eq:resVarGraphene}. For the coupling
through gauge potential {[}$V_{2}$ matrix elements in Eq.~\eqref{eq:potential}{]}
the parametric difference between monolayer and bilayer amounts to
the replacement $v_{F}^{2}\rightarrow\hbar^{2}k_{F}^{2}/(2m)^{2}$
(after taking the square of the matrix elements and using the quasielastic
approximation). When multiplied by the square of the density of states
in bilayer graphene, $\mathcal{D}(E)\sim m/\hbar^{2}$, we obtain
the factor $k_{F}^{2}/\hbar^{2}$, which has exactly the same parametric
dependence appearing for single layer graphene -- there, the factor
 $v_{F}^{2}$
multiplies the square of the density of states of the monolayer, $\mathcal{D}(E)\sim k/(\hbar v_{F})$.

Despite qualitative similarities there are apparent quantitative differences.
A striking one is the overall suppression of resistivity in bilayer
graphene, which is clearly seen when we compare Fig.~\ref{fig:res}(a)
and~\ref{fig:res}(b). This is due to the higher stiffness and mass
density of bilayer graphene (see Sec.~\ref{sec:phonons}), and to
the $1/2$ term in Eq.~\eqref{eq:H2_bi} for the gauge potential
which, at odds with the parametric dependence just discussed, does
not cancel out in the expression for the resistivity. Also, the scalar
potential contribution is quantitative different, being enhanced in
bilayer graphene, as can be seen by comparing Eqs.~\eqref{eq:resLowTIn},
and~\eqref{eq:resHighTIn}, Eqs.~\eqref{eq:resNsFlexLowT} and~\eqref{eq:resNsFlexHighT},
and Eqs.~\eqref{eq:resFlexStrLowT} and~\eqref{eq:resFlexStrHighT}
with their monolayer counterparts. This arises because pseudo-spin
conservation allows back scattering due to scalar potential in bilayer
but not in monolayer graphene.

The quantitative discrepancy between the $T$ dependent resistivity
in bilayer and monolayer graphene originates an interesting difference
regarding room $T$ mobility in non-strained samples. The mobility
$\mu$, defined as $\varrho=1/(en\mu)$, is in the non-strained case
limited by PF scattering (see Sec.~\ref{sub:resNstrFP}) and takes
the form \begin{equation}
\mu\approx\mathcal{A}\frac{64\pi\hbar ev_{F}^{2}}{k_{B}^{2}T^{2}},\label{eq:mobility}\end{equation}
with $\mathcal{A}_{B}=\kappa^{2}/[g^{2}+\hbar^{2}v_{F}^{2}\beta^{2}/(8a^{2})]$
in bilayer graphene and $\mathcal{A}_{M}=\kappa^{2}/[g^{2}/2+\hbar^{2}v_{F}^{2}\beta^{2}/(4a^{2})]$
in monolayer graphene, ignoring the logarithmic contribution of order
unity, Eq.~\eqref{eq:resNsFlexHighT}. For monolayer graphene at
room $T$ the mobility is limited to the value for samples on substrate,
$\mu\approx198\mathcal{A}_{M}\sim1\,\,\mbox{m}^{2}/\mbox{Vs}$, as
has recently been confirmed experimentally.\cite{COK+10} For bilayer
graphene, however, the quantitative differences discussed above lead
to an enhanced room $T$ mobility, $\mu\approx198\mathcal{A}_{B}\sim20\,\,\mbox{m}^{2}/\mbox{Vs}$.
This might be an interesting aspect to take into account regarding
room $T$ electronic applications. Reports of much smaller mobility
(one order of magnitude) in recent experiments in suspended bilayer
graphene\cite{FMY09} might be an indication that residual, $T$ independent
scattering is at work, overcoming the intrinsic FP contribution.

Another remark worth discussion is the validity of results in the
non-strained regime, in particular Eq.~\eqref{eq:resNsFlexHighT}
for the high $T$ resistivity. At low densities, when $k_{F}$ becomes
comparable with the infrared cutoff $q_{c}$ given by the onset of
anharmonic effects,\cite{ZRF+10} the harmonic approximation used
here breaks down. A complete theory would require taking into account
anharmonicities, but this is beyond the scope of the present work.
Nevertheless, it is likely that unavoidable little strain $u$ is
always present in real samples,\cite{COK+10} and this increases the
validity of the harmonic approximation.\cite{RFZK11} Moreover, the infrared cutoff $q_{c}$ due to anharmonic effects depends
on the applied strain (see Appendix~\ref{secapp:anharmonic}), decreasing
as strain increases. This is consistent with sample to sample mobility
differences of order unity recently reported in suspended monolayer
graphene,\cite{COK+10} where strain $u\lesssim10^{-4}-10^{-3}$ is
naturally expected.

Finally, we comment on a recent theory paper by Mariani and von~Oppen\cite{MvO10}
where the $T$ dependent resistivity of monolayer graphene has been
fully discussed. In the high $T$ regime, $T\gg T_{BG}$, our results
for the monolayer case agree with those of Ref.~\onlinecite{MvO10}.
However, at low $T$, i.e. $T\ll T_{BG}$, the authors of Ref.~\onlinecite{MvO10}
found a new regime where the scalar potential contribution dominates.
This happens because in Ref.~\onlinecite{MvO10} the electron-phonon
coupling from scalar potential is assumed to be much higher than the
gauge potential coupling, unless $T\ll T_{GD}$, where $T_{GD}$ is
the energy scale at which screening becomes relevant and scalar and
gauge potentials become comparable. The new regime arises for $T_{GD}\ll T\ll T_{BG}$.
With the parameter values used in the present work, however, the electron-phonon
coupling due to scalar potential is always similar or smaller than
the gauge potential (see Sec.~\ref{sub:coupling}). Therefore, we
can neglect this low $T$ contribution since it gives higher power
law behavior than the gauge potential. Recent $T$ dependent resistivity
due to in-plane phonons measured in single layer graphene at the high
densities\cite{EK10} seem to corroborate the latter picture.

\section{Conclusions}

\label{sec:Conclusions}

In the present work we have studied the $T$ dependent resistivity
due to scattering by both acoustic in-plane phonons and FPs in doped,
suspended bilayer graphene. We have found the bilayer membrane to
follow the qualitative behavior of the monolayer cousin.\cite{MvO10,COK+10}
Different electronic structure combine with different electron-phonon
coupling to give the same parametric dependence in resistivity, and
in particular the same $T$ behavior. In parallel with the single
layer, FPs dominate the phonon contribution to resistivity in the
absence of strain, where a density independent mobility is obtained.
This contribution is strongly suppressed by tension, similarly to
monolayer graphene.\cite{COK+10} However, an interesting quantitative
difference with respect to suspended monolayer has been found. In
the latter, as shown in Ref.~\onlinecite{COK+10}, FPs limit room
$T$ mobility $\mu$ to values obtained for samples on substrate,
$\mu\sim1\,\,\mbox{m\ensuremath{^{2}/}(Vs)}$, when tension is absent.
In bilayer, quantitative differences in electron-phonon coupling and
elastic constants lead to a room $T$ mobility enhanced by one order
of magnitude, $\mu\sim20\,\,\mbox{m\ensuremath{^{2}/}(Vs)}$, even
in non-strained samples. This finding has obvious advantages for room
$T$ electronic applications. It has also been shown that for a correct
description of acoustic phonon scattering in both monolayer and bilayer
graphene, even at the qualitative level, coupling to both scalar and
gauge potentials needs to be taken into account.

\section*{Acknowledgments}

We acknowledge financial support from
MICINN (Spain) through grants FIS2008-00124 and CONSOLIDER
CSD2007-00010, and from the Comunidad de Madrid, through
NANOBIOMAG. E.V.C. acknowledges
financial support from the Juan de la Cierva Program (MICINN, Spain),
and the Program {}``Est\'imulo \`a
Investigaç\~ao'' of Gulbenkian Foundation, Portugal.
M.I.K. acknowledges a financial support from FOM (The
Netherlands).

\appendix

\section{Collision integral}

\label{sec:APPcollInt}

The rate of change of $f_{\mathbf{k}}$ due to scattering, the so-called
collision integral $\dot{f}_{\mathbf{k}}\bigr|_{\mathrm{scatt}}$
appearing on the right hand side of the Boltzmann equation Eq.~\eqref{eq:BoltzEq},
is the difference between the rate at which quasiparticles enter the
state $\bigl|\mathbf{k}\bigr\rangle$ and the rate at which they leave
it,\begin{equation}
\dot{f}_{\mathbf{k}}\bigr|_{\mathrm{scatt}}=\sum_{\mathbf{k}'}\left[f_{\mathbf{k}'}(1-f_{\mathbf{k}})\mathcal{W}_{\mathbf{k}'}^{\mathbf{k}}-f_{\mathbf{k}}(1-f_{\mathbf{k}'})\mathcal{W}_{\mathbf{k}}^{\mathbf{k}'}\right],\label{eq:fDotScatt}\end{equation}
where $\mathcal{W}_{i}^{f}$ is the scattering probability between
state $\bigl|i\bigr\rangle$ and $\bigl|f\bigr\rangle$. Here we use
Fermi's golden rule, which reads\begin{equation}
\mathcal{W}_{i}^{f}=\frac{2\pi}{\hbar}\Bigl|\bigl\langle f\bigr|\mathcal{H}_{int}\bigl|i\bigr\rangle\Bigr|^{2}\delta(\mathcal{E}_{f}-\mathcal{E}_{i}),\label{eq:Golden}\end{equation}
and is equivalent to rest upon the Born approximation for the differential
scattering cross-section. In this appendix we provide explicit expressions
for the integral collision Eq.~\eqref{eq:fDotScatt} arising due
to scattering by phonons in bilayer graphene (and also monolayer for
comparison).

The crucial step to get $\dot{f}_{\mathbf{k}}\bigr|_{\mathrm{scatt}}$
is finding the scattering probability for a quasi-particle in state
$\left|\mathbf{k}\right\rangle $ to be scattered into state $\left|\mathbf{k}'\right\rangle $,
i.e. $\mathcal{W}_{\mathbf{k}}^{\mathbf{k}'}$ (since the process
is quasi-elastic interband transitions are not allowed, meaning that
both states belong to the same band). The scattering mechanism is
encoded in the interaction $\mathcal{H}_{int}$, which in the present
case is given by $H_{ep}$ in Eq.~\eqref{eq:HepCompact}. It is readily
seen that scattering occurs only through emission or absorption of
one phonon or emission/absorption of two phonons. The initial and
final states are thus tensorial products of the form $\left|i\right\rangle =\left|\mathbf{k}\right\rangle \otimes\left|n_{\mathbf{q}}\right\rangle $
or $\left|i\right\rangle =\left|\mathbf{k}\right\rangle \otimes\left|n_{\mathbf{q}},n_{\mathbf{q'}}\right\rangle $,
and $\left|f\right\rangle =\left|\mathbf{k}'\right\rangle \otimes\left|n_{\mathbf{q}}\pm1\right\rangle $,
$\left|f\right\rangle =\left|\mathbf{k}'\right\rangle \otimes\left|n_{\mathbf{q}}\pm1,n_{\mathbf{q}'}\pm1\right\rangle $
or $\left|f\right\rangle =\left|\mathbf{k}'\right\rangle \otimes\left|n_{\mathbf{q}}\pm1,n_{\mathbf{q}'}^{F}\mp1\right\rangle $,
where $\left|n_{\mathbf{q}}\right\rangle $ and $\left|n_{\mathbf{q}},n_{\mathbf{q'}}\right\rangle $
represent one and two phonon states in the occupation number representation,\cite{MO08}
and the electron like quasiparticle state is written according to
the unitary transformation in Eq.~\eqref{eq:rot} as $\left|\mathbf{k}\right\rangle =(e^{-i\theta_{\mathbf{k}}}a_{\mathbf{k}}^{\dagger}\left|0\right\rangle +e^{i\theta_{\mathbf{k}}}b_{\mathbf{k}}^{\dagger}\left|0\right\rangle )/\sqrt{2}$
(electron-hole symmetry guarantees the result is the same for both
electron and hole doping).

In order to obtain $\left|\left\langle f\right|H_{ep}\left|i\right\rangle \right|^{2}$,
with $\left|i\right\rangle $ and $\left|f\right\rangle $ as given
above, we take the following steps. (i)~Terms of the form $V_{1}V_{2}$,
where $V_{1}$ stands for scalar potential and $V_{2}$ for gauge
potential induced matrix elements in Eq.~\eqref{eq:potential}, are
neglected. It is easy to show that such terms come proportional to
oscillatory factors $e^{\pm i2\theta_{\mathbf{k}}}$ or $e^{\pm i2\theta_{\mathbf{k}'}}$
(in monolayer graphene $e^{\pm i\theta_{\mathbf{k}}}$ or $e^{\pm i\theta_{\mathbf{k}'}}$),
stemming from the unitary transformation in Eq.~\eqref{eq:rot}.
These terms can safely be neglected in doing the summation over the
direction of $\mathbf{k}$ and $\mathbf{k}'$ in the numerator of
Eq.~\eqref{eq:resVarGraphene}, keeping $\theta_{\mathbf{k},\mathbf{k}'}$
fixed. The resistivity is then the sum of two independent contributions,
originating from scalar and gauge potentials, well in the spirit of
Matthiessen's empirical rule.\cite{zimanEP} (ii)~For the scalar
potential contribution the terms $\left|V_{1,\mathbf{q}}^{\nu}e^{i(\theta_{\mathbf{k}'}-\theta_{\mathbf{k}})}+V_{1,\mathbf{q}}^{\nu}e^{i(\theta_{\mathbf{k}'}-\theta_{\mathbf{k}})}\right|^{2}/4$
are proportional to the overlap of states belonging to the same band
$(1+\cos2\theta_{\mathbf{k},\mathbf{k}'})/2$, and can be written
as \begin{equation}
\left|V_{1,\mathbf{q}}^{\nu}\frac{e^{i(\theta_{\mathbf{k}'}-\theta_{\mathbf{k}})}}{2}+V_{1,\mathbf{q}}^{\nu}\frac{e^{-i(\theta_{\mathbf{k}'}-\theta_{\mathbf{k}})}}{2}\right|^{2}=\left|V_{1,\mathbf{q}}^{\nu}\right|^{2}\frac{1+\cos2\theta_{\mathbf{k},\mathbf{k}'}}{2}.\label{eq:modsq1}\end{equation}
The same manipulation holds for two phonon terms, with $V_{1,\mathbf{q}}^{\nu}\rightarrow V_{1,\mathbf{q},\mathbf{q}'}^{F}$.
(iii)~For the gauge potential contribution there are oscillatory
terms which, owing to the argument of point (i), can be neglected,\begin{multline}
\left|V_{2,\mathbf{q},\mathbf{k},\mathbf{k}'}^{\nu}\frac{e^{i(\theta_{\mathbf{k}'}+\theta_{\mathbf{k}})}}{2}+\Bigl(V_{2,-\mathbf{q},\mathbf{k},\mathbf{k}'}^{\nu}\Bigr)^{*}\frac{e^{-i(\theta_{\mathbf{k}'}+\theta_{\mathbf{k}})}}{2}\right|^{2}\simeq\\
\left|\tilde{V}_{2,\mathbf{q}}^{\nu}\right|^{2}\left(\frac{k^{2}}{2}+\frac{k'^{2}}{2}+kk'\cos\theta_{\mathbf{k},\mathbf{k}'}\right),\label{eq:modsq2}\end{multline}
where we used $\tilde{V}$ to express the matrix elements given in
Eq.~\eqref{eq:potential} for bilayer graphene without the term $(\pi_{\mathbf{k}}+\pi_{\mathbf{k}'})$.
A similar manipulation holds for two phonon terms, with $\tilde{V}_{2,\mathbf{q}}^{\nu}\rightarrow\tilde{V}_{2,\mathbf{q},\mathbf{q}'}^{F}$.

Finally, summing over phonon momenta and doing the thermal average,
we can write $\mathcal{W}_{\mathbf{k}}^{\mathbf{k}'}$ as follows:
when scattering is via one phonon, \begin{align}
\mathcal{W}_{\mathbf{k}}^{\mathbf{k}'} & =\frac{\pi}{\hbar}\sum_{\mathbf{q}}w_{\nu}(\mathbf{q},\mathbf{k},\mathbf{k}')n_{\mathbf{q}}\delta_{\mathbf{k}',\mathbf{k}+\mathbf{q}}\delta(\varepsilon_{\mathbf{k}'}-\varepsilon_{\mathbf{k}}-\hbar\omega_{\mathbf{q}}^{\nu})\nonumber \\
+ & \frac{\pi}{\hbar}\sum_{\mathbf{q}}w_{\nu}(\mathbf{q},\mathbf{k},\mathbf{k}')(n_{\mathbf{q}}+1)\delta_{\mathbf{k}',\mathbf{k}-\mathbf{q}}\delta(\varepsilon_{\mathbf{k}'}-\varepsilon_{\mathbf{k}}+\hbar\omega_{\mathbf{q}}^{\nu}),\label{eq:Wkkl1}\end{align}
where the first term is due to absorption and the second to emission
of a single phonon; when scattering involves two phonons, \begin{align}
\mathcal{W}_{\mathbf{k}}^{\mathbf{k}'} & =\frac{\pi}{\hbar}\sum_{\mathbf{q},\mathbf{q}'}w_{F}(\mathbf{q},\mathbf{q}',\mathbf{k},\mathbf{k}')n_{\mathbf{q}}n_{\mathbf{q}'}\times\nonumber \\
 & \delta_{\mathbf{k}',\mathbf{k}+\mathbf{q}+\mathbf{q}'}\delta(\varepsilon_{\mathbf{k}'}-\varepsilon_{\mathbf{k}}-\hbar\omega_{\mathbf{q}}^{F}-\hbar\omega_{\mathbf{q}'}^{F})\nonumber \\
+ & \frac{\pi}{\hbar}\sum_{\mathbf{q},\mathbf{q}'}w_{F}(\mathbf{q},\mathbf{q}',\mathbf{k},\mathbf{k}')(n_{\mathbf{q}}+1)(n_{\mathbf{q}'}+1)\times\nonumber \\
 & \delta_{\mathbf{k}',\mathbf{k}-\mathbf{q}-\mathbf{q}'}\delta(\epsilon_{\mathbf{k}'}-\epsilon_{\mathbf{k}}+\hbar\omega_{\mathbf{q}}^{F}+\hbar\omega_{\mathbf{q}'}^{F})\nonumber \\
+ & \frac{2\pi}{\hbar}\sum_{\mathbf{q},\mathbf{q}'}w_{F}(\mathbf{q},\mathbf{q}',\mathbf{k},\mathbf{k}')(n_{\mathbf{q}}+1)n_{\mathbf{q}'}\times\nonumber \\
 & \delta_{\mathbf{k}',\mathbf{k}-\mathbf{q}+\mathbf{q}'}\delta(\epsilon_{\mathbf{k}'}-\epsilon_{\mathbf{k}}+\hbar\omega_{\mathbf{q}}^{F}-\hbar\omega_{\mathbf{q}'}^{F}),\label{eq:Wkkl2}\end{align}
where the first term is due to absorption of two FPs, the second to
emission of two FPs, and the last one comes from absorption of a single
FP and emission of another one. The kernels $w_{\nu}(\mathbf{q},\mathbf{k},\mathbf{k}')$
and $w_{F}(\mathbf{q},\mathbf{q}',\mathbf{k},\mathbf{k}')$ represent
the sum of the right hand side of Eq.~\eqref{eq:modsq1} with Eq.~\eqref{eq:modsq2},
as given in Eq.~\eqref{eq:wKernelLT2L}. For Monolayer graphene $\mathcal{W}_{\mathbf{k}}^{\mathbf{k}'}$
take exactly the same form;\cite{MvO10} only the kernels change,
being given instead by Eq.~\eqref{eq:wKernelLT1L}. The collision
integral may finally be put in the form,\begin{align}
\dot{f}_{\mathbf{k}}\bigr|_{\mathrm{scatt}} & =\frac{\pi}{\hbar}\sum_{\mathbf{k}'}\sum_{\mathbf{q},\nu}w_{\nu}(\mathbf{q},\mathbf{k},\mathbf{k}')\times\nonumber \\
\times & \left\{ \left[f_{\mathbf{k}'}(1-f_{\mathbf{k}})n_{\mathbf{q}}-f_{\mathbf{k}}(1-f_{\mathbf{k}'})(n_{\mathbf{q}}+1)\right]\right.\nonumber \\
 & \delta_{\mathbf{k},\mathbf{k}'+\mathbf{q}}\delta(\varepsilon_{\mathbf{k}}-\varepsilon_{\mathbf{k}'}-\hbar\omega_{\mathbf{q}}^{\nu})\nonumber \\
+ & \left[f_{\mathbf{k}'}(1-f_{\mathbf{k}})(n_{\mathbf{q}}+1)-f_{\mathbf{k}}(1-f_{\mathbf{k}'})n_{\mathbf{q}}\right]\nonumber \\
 & \left.\delta_{\mathbf{k},\mathbf{k}'-\mathbf{q}}\delta(\varepsilon_{\mathbf{k}}-\varepsilon_{\mathbf{k}'}+\hbar\omega_{\mathbf{q}}^{\nu})\right\} ,\label{eq:collInt1ph}\end{align}
for one phonon scattering processes, and\begin{align}
\dot{f}_{\mathbf{k}}\bigr|_{\mathrm{scatt}} & =\frac{\pi}{\hbar}\sum_{\mathbf{k}'}\sum_{\mathbf{q},\mathbf{q}'}w_{F}(\mathbf{q},\mathbf{q}',\mathbf{k},\mathbf{k}')\times\nonumber \\
\times & \left\{ \left[f_{\mathbf{k}'}(1-f_{\mathbf{k}})(n_{\mathbf{q}}+1)(n_{\mathbf{q}'}+1)-f_{\mathbf{k}}(1-f_{\mathbf{k}'})n_{\mathbf{q}}n_{\mathbf{q}'}\right]\right.\nonumber \\
 & \delta_{\mathbf{k},\mathbf{k}'-\mathbf{q}-\mathbf{q}'}\delta(\varepsilon_{\mathbf{k}}-\varepsilon_{\mathbf{k}'}+\hbar\omega_{\mathbf{q}}^{F}+\hbar\omega_{\mathbf{q}'}^{F})\nonumber \\
+ & \left[f_{\mathbf{k}'}(1-f_{\mathbf{k}})(n_{\mathbf{q}}+1)n_{\mathbf{q}'}-f_{\mathbf{k}}(1-f_{\mathbf{k}'})n_{\mathbf{q}}(n_{\mathbf{q}'}+1)\right]\nonumber \\
 & \delta_{\mathbf{k},\mathbf{k}'-\mathbf{q}+\mathbf{q}'}\delta(\varepsilon_{\mathbf{k}}-\varepsilon_{\mathbf{k}'}+\hbar\omega_{\mathbf{q}}^{F}-\hbar\omega_{\mathbf{q}'}^{F})\nonumber \\
+ & \left[f_{\mathbf{k}'}(1-f_{\mathbf{k}})n_{\mathbf{q}}n_{\mathbf{q}'}-f_{\mathbf{k}}(1-f_{\mathbf{k}'})(n_{\mathbf{q}}+1)(n_{\mathbf{q}'}+1)\right]\nonumber \\
 & \delta_{\mathbf{k},\mathbf{k}'+\mathbf{q}+\mathbf{q}'}\delta(\varepsilon_{\mathbf{k}}-\varepsilon_{\mathbf{k}'}-\hbar\omega_{\mathbf{q}}^{F}-\hbar\omega_{\mathbf{q}'}^{F})\nonumber \\
+ & \left[f_{\mathbf{k}'}(1-f_{\mathbf{k}})n_{\mathbf{q}}(n_{\mathbf{q}'}+1)-f_{\mathbf{k}}(1-f_{\mathbf{k}'})(n_{\mathbf{q}}+1)n_{\mathbf{q}'}\right]\nonumber \\
 & \left.\delta_{\mathbf{k},\mathbf{k}'+\mathbf{q}-\mathbf{q}'}\delta(\varepsilon_{\mathbf{k}}-\varepsilon_{\mathbf{k}'}-\hbar\omega_{\mathbf{q}}^{F}+\hbar\omega_{\mathbf{q}'}^{F})\right\} ,\label{eq:collInt2ph}\end{align}
for scattering through two FPs.

\section{Linearized collision integral}

\label{sec:APPlinCollInt}

In this appendix we derive the linearized version of the collision
integrals given in Eqs.~\eqref{eq:collInt1ph} and~\eqref{eq:collInt2ph}.
We start by expanding electron and phonon probability distributions
around their equilibrium values,\begin{equation}
f_{\mathbf{k}}=f_{\mathbf{k}}^{(0)}+\delta f_{\mathbf{k}},\hspace{1em}\hspace{1em}n_{\mathbf{q}}=n_{\mathbf{q}}^{(0)}+\delta n_{\mathbf{q}},\label{eq:linAPP}\end{equation}
where the variations can be written as $\delta f_{\mathbf{k}}=-\frac{\partial f_{\mathbf{k}}^{(0)}}{\partial\varepsilon_{\mathbf{k}}}\varphi_{\mathbf{k}}$
{[}see Eq.~\eqref{eq:lin}{]} and $\delta n_{\mathbf{q}}=-\frac{\partial n_{\mathbf{q}}^{(0)}}{\partial(\hbar\omega_{\mathbf{q}})}\chi_{\mathbf{q}}$.
The linearized collision integral $\delta\dot{f}_{\mathbf{k}}\bigr|_{\mathrm{scatt}}$
is then obtained by expanding $\dot{f}_{\mathbf{k}}\bigr|_{\mathrm{scatt}}$
up to first order in the variations.\cite{zimanEP,LL10}

\subsection{One phonon scattering}

This case follows closely the steps outlined in Ref.~\onlinecite{LL10},
and for the case of monolayer graphene it has been derived in Ref.~\onlinecite{MvO10}.
Since the difference between monolayer and bilayer amounts to a different
kernel $w_{\nu}(\mathbf{q},\mathbf{k},\mathbf{k}')$ in Eqs.~\eqref{eq:collInt1ph},
which does not play any role in the linearization, we can directly
apply the result of Ref.~\onlinecite{MvO10} to the present case.
In order to set notation for the more elaborated case of two phonon
scattering, we outline the main steps of the derivation in the following.

We first note that at equilibrium detailed balance implies $\dot{f}_{\mathbf{k}}^{(0)}\bigr|_{\mathrm{scatt}}=0$,
from which we get the relation\begin{equation}
f_{\mathbf{k}'}^{(0)}(1-f_{\mathbf{k}}^{(0)})n_{\mathbf{q}}^{(0)}=f_{\mathbf{k}}^{(0)}(1-f_{\mathbf{k}'}^{(0)})(n_{\mathbf{q}}^{(0)}+1),\label{eq:equilRel}\end{equation}
which can be easily verified by direct calculation.\cite{LL10} Therefore,
in order to get the linearized collision integral it is enough to
calculate the variation,\begin{multline}
\delta\left[f_{\mathbf{k}'}(1-f_{\mathbf{k}})n_{\mathbf{q}}-f_{\mathbf{k}}(1-f_{\mathbf{k}'})(n_{\mathbf{q}}+1)\right]=\\
(1-f_{\mathbf{k}}^{(0)})(1-f_{\mathbf{k}'}^{(0)})(n_{\mathbf{q}}^{(0)}+1)\delta\left(\frac{f_{\mathbf{k}'}}{1-f_{\mathbf{k}'}}\frac{n_{\mathbf{q}}}{n_{\mathbf{q}}+1}-\frac{f_{\mathbf{k}}}{1-f_{\mathbf{k}}}\right).\label{eq:var}\end{multline}
The variations appearing on the right hand side of Eq.~\eqref{eq:var}
can be computed easily by noting that \begin{equation}
\delta\left(\frac{f}{1-f}\right)=\frac{\delta f}{(1-f^{(0)})^{2}}\hspace{1em}\mbox{and}\hspace{1em}\delta\left(\frac{f}{n+1}\right)=\frac{\delta n}{(n^{(0)}+1)^{2}}.\label{eq:deltaFrac}\end{equation}
Using Eq.~\eqref{eq:equilRel}, and rewriting $\delta f_{\mathbf{k}}$
and $\delta n_{\mathbf{q}}$ as\begin{equation}
\delta f_{\mathbf{k}}=f_{\mathbf{k}}^{(0)}(1-f_{\mathbf{k}}^{(0)})\frac{\varphi_{\mathbf{k}}}{k_{B}T}\hspace{1em}\delta n_{\mathbf{q}}=n_{\mathbf{q}}^{(0)}(n_{\mathbf{q}}^{(0)}+1)\frac{\chi_{\mathbf{q}}}{k_{B}T},\label{eq:linDerivAPP}\end{equation}
it is easy to show that\begin{multline}
\delta\left[f_{\mathbf{k}'}(1-f_{\mathbf{k}})n_{\mathbf{q}}-f_{\mathbf{k}}(1-f_{\mathbf{k}'})(n_{\mathbf{q}}+1)\right]=\\
\frac{1}{k_{B}T}f_{\mathbf{k}}^{(0)}(1-f_{\mathbf{k}'}^{(0)})(n_{\mathbf{q}}^{(0)}+1)\left(\varphi_{\mathbf{k}'}+\chi_{\mathbf{q}}-\varphi_{\mathbf{k}}\right).\end{multline}
The linearized collision integral may then be obtained,\begin{widetext}\begin{eqnarray}
\delta\dot{f}_{\mathbf{k}}\bigr|_{\mathrm{scatt}} & = & -\frac{\pi}{\hbar k_{B}T}\sum_{\mathbf{k}'}\sum_{\mathbf{q},\nu}w_{\nu}(\mathbf{q},\mathbf{k},\mathbf{k}')f_{\mathbf{k}}^{(0)}(1-f_{\mathbf{k}'}^{(0)})(n_{\mathbf{q}}^{(0)}+1)\left(\varphi_{\mathbf{k}'}+\chi_{\mathbf{q}}-\varphi_{\mathbf{k}}\right)\delta_{\mathbf{k},\mathbf{k}'+\mathbf{q}}\delta(\varepsilon_{\mathbf{k}}-\varepsilon_{\mathbf{k}'}-\hbar\omega_{\mathbf{q}}^{\nu})\nonumber \\
 & - & \frac{\pi}{\hbar k_{B}T}\sum_{\mathbf{k}'}\sum_{\mathbf{q},\nu}w_{\nu}(\mathbf{q},\mathbf{k},\mathbf{k}')(1-f_{\mathbf{k}}^{(0)})f_{\mathbf{k}'}^{(0)}(n_{\mathbf{q}}^{(0)}+1)\left(\varphi_{\mathbf{k}'}-\varphi_{\mathbf{k}}-\chi_{\mathbf{q}}\right)\delta_{\mathbf{k},\mathbf{k}'-\mathbf{q}}\delta(\varepsilon_{\mathbf{k}}-\varepsilon_{\mathbf{k}'}+\hbar\omega_{\mathbf{q}}^{\nu}).\label{eq:linColIntTot1ph}\end{eqnarray}
\end{widetext}

Now we introduce two typical approximations: consider phonons at equilibrium
by taking $\chi_{\mathbf{q}}\approx0$, so that $n_{\mathbf{q}}\approx n_{\mathbf{q}}^{(0)}$,
valid at not too low temperatures;\cite{zimanEP} consider quasielastic
scattering, with $\varepsilon_{\mathbf{k}},\varepsilon_{\mathbf{k}'}\gg\hbar\omega_{\mathbf{q}}$.
The latter approximation enables us to rewrite the delta functions,
$\delta(\varepsilon_{\mathbf{k}}-\varepsilon_{\mathbf{k}'}\pm\hbar\omega_{\mathbf{q}}^{\nu})\rightarrow\delta(\varepsilon_{\mathbf{k}}-\varepsilon_{\mathbf{k}'})$,
and to approximate $f_{\mathbf{k}}^{(0)}(1-f_{\mathbf{k}'}^{(0)})$
by\cite{LL10}\begin{equation}
f_{\mathbf{k}}^{(0)}(1-f_{\mathbf{k}'}^{(0)})=(f_{\mathbf{k}'}^{(0)}-f_{\mathbf{k}}^{(0)})n_{\mathbf{q}}^{(0)}\approx\pm\hbar\omega_{\mathbf{q}}\frac{\partial f_{\mathbf{k}}^{(0)}}{\partial\epsilon_{\mathbf{k}}}n_{\mathbf{q}}^{(0)}.\label{eq:quasielast}\end{equation}
The linearized collision integral then reads\begin{multline}
\delta\dot{f}_{\mathbf{k}}\bigr|_{\mathrm{scatt}}=-\frac{2\pi}{\hbar}\sum_{\mathbf{k}'}\sum_{\mathbf{q},\nu}w_{\nu}(\mathbf{q},\mathbf{k},\mathbf{k}')\omega_{\mathbf{q}}^{\nu}\times\\
\frac{\partial n_{\mathbf{q}}}{\partial\omega_{\mathbf{q}}^{\nu}}\frac{\partial f_{\mathbf{k}}^{(0)}}{\partial\epsilon_{\mathbf{k}}}(\varphi_{\mathbf{k}}-\varphi_{\mathbf{k}'})\delta_{\mathbf{k},\mathbf{k}'+\mathbf{q}}\delta(\epsilon_{\mathbf{k}}-\epsilon_{\mathbf{k}'}),\label{eq:linColInt1ph}\end{multline}
where we have used equalities $w_{\nu}(\mathbf{q},\mathbf{k},\mathbf{k}')=w_{\nu}(-\mathbf{q},\mathbf{k},\mathbf{k}')$
and $\omega_{\mathbf{q}}=\omega_{-\mathbf{q}}$. Finally, Eq.~\eqref{eq:linColInt1ph}
can be put in the form of Eq.~\eqref{eq:linColInt},\begin{equation}
\dot{f}_{\mathbf{k}}\bigr|_{\mathrm{scatt}}=-\sum_{\mathbf{k}'}\mathcal{P}_{\mathbf{k},\mathbf{k}'}(\varphi_{\mathbf{k}}-\varphi_{\mathbf{k}'}),\label{eq:linColIntQE}\end{equation}
where $\mathcal{P}_{\mathbf{k},\mathbf{k}'}$ is given in Eq.~\eqref{eq:P1ph}.

\subsection{Two phonon scattering}

Now we proceed with the linearization of the collision integral in
Eq.~\eqref{eq:collInt2ph}, originating from scattering processes
involving two FPs. At equilibrium detailed balance is guaranteed,
$\dot{f}_{\mathbf{k}}^{(0)}\bigr|_{\mathrm{scatt}}=0$, and the following
two relations hold,\begin{eqnarray}
\frac{f_{\mathbf{k}'}^{(0)}}{1-f_{\mathbf{k}'}^{(0)}} & = & \frac{f_{\mathbf{k}}^{(0)}}{1-f_{\mathbf{k}}^{(0)}}\frac{n_{\mathbf{q}}^{(0)}}{n_{\mathbf{q}}^{(0)}+1}\frac{n_{\mathbf{q}'}^{(0)}}{n_{\mathbf{q}'}+1},\nonumber \\
\frac{f_{\mathbf{k}'}^{(0)}}{1-f_{\mathbf{k}'}^{(0)}}\frac{n_{\mathbf{q}'}^{(0)}}{n_{\mathbf{q}'}^{(0)}+1} & = & \frac{f_{\mathbf{k}}^{(0)}}{1-f_{\mathbf{k}}^{(0)}}\frac{n_{\mathbf{q}}^{(0)}}{n_{\mathbf{q}}^{(0)}+1}.\label{eq:detbal2ph}\end{eqnarray}
In order to get the linearized collision integral it is easy to see
that we only need the following two variations,

\begin{multline}
\delta\left[f_{\mathbf{k}'}(1-f_{\mathbf{k}})(n_{\mathbf{q}}+1)(n_{\mathbf{q}'}+1)-f_{\mathbf{k}}(1-f_{\mathbf{k}'})n_{\mathbf{q}}n_{\mathbf{q}'}\right]=\\
(1-f_{\mathbf{k}}^{(0)})(1-f_{\mathbf{k}'}^{(0)})(n_{\mathbf{q}}^{(0)}+1)(n_{\mathbf{q}'}^{(0)}+1)\times\\
\delta\left(\frac{f_{\mathbf{k}'}}{1-f_{\mathbf{k}'}}-\frac{f_{\mathbf{k}}}{1-f_{\mathbf{k}}}\frac{n_{\mathbf{q}}}{n_{\mathbf{q}}+1}\frac{n_{\mathbf{q}'}}{n_{\mathbf{q}'}+1}\right),\label{eq:var1a}\end{multline}
and\begin{multline}
\delta\left[f_{\mathbf{k}'}(1-f_{\mathbf{k}})(n_{\mathbf{q}}+1)n_{\mathbf{q}'}-f_{\mathbf{k}}(1-f_{\mathbf{k}'})n_{\mathbf{q}}(n_{\mathbf{q}'}+1)\right]=\\
(1-f_{\mathbf{k}}^{(0)})(1-f_{\mathbf{k}'}^{(0)})(n_{\mathbf{q}}^{(0)}+1)(n_{\mathbf{q}'}^{(0)}+1)\times\\
\delta\left(\frac{f_{\mathbf{k}'}}{1-f_{\mathbf{k}'}}\frac{n_{\mathbf{q}'}}{n_{\mathbf{q}'}+1}-\frac{f_{\mathbf{k}}}{1-f_{\mathbf{k}}}\frac{n_{\mathbf{q}}}{n_{\mathbf{q}}+1}\right),\label{eq:var2a}\end{multline}
the other two possibilities being related with these ones by a minus
sign and $\mathbf{k}\rightarrow\mathbf{k}'$. The variations appearing
on the right hand side of Eqs.~\eqref{eq:var1a} and~\eqref{eq:var2a}
can be computed easily by using Eq.~\eqref{eq:deltaFrac}. We then
arrive at the variations\begin{multline}
\delta\left[f_{\mathbf{k}'}(1-f_{\mathbf{k}})(n_{\mathbf{q}}+1)(n_{\mathbf{q}'}+1)-f_{\mathbf{k}}(1-f_{\mathbf{k}'})n_{\mathbf{q}}n_{\mathbf{q}'}\right]=\\
(1-f_{\mathbf{k}}^{(0)})f_{\mathbf{k}'}^{(0)}(n_{\mathbf{q}}^{(0)}+1)(n_{\mathbf{q}'}^{(0)}+1)\frac{\left(\varphi_{\mathbf{k}'}-\varphi_{\mathbf{k}}-\chi_{\mathbf{q}}-\chi_{\mathbf{q}'}\right)}{k_{B}T}\label{eq:var1b}\end{multline}
and\begin{multline}
\delta\left[f_{\mathbf{k}'}(1-f_{\mathbf{k}})(n_{\mathbf{q}}+1)n_{\mathbf{q}'}-f_{\mathbf{k}}(1-f_{\mathbf{k}'})n_{\mathbf{q}}(n_{\mathbf{q}'}+1)\right]=\\
(1-f_{\mathbf{k}}^{(0)})f_{\mathbf{k}'}^{(0)}(n_{\mathbf{q}}^{(0)}+1)n_{\mathbf{q}'}^{(0)}\frac{\left(\varphi_{\mathbf{k}'}+\chi_{\mathbf{q}'}-\varphi_{\mathbf{k}}-\chi_{\mathbf{q}}\right)}{k_{B}T},\label{eq:var2b}\end{multline}
where we used Eq.~\eqref{eq:linDerivAPP} and the relations in Eq.~\eqref{eq:detbal2ph}.
It is convenient to express the quantity $(1-f_{\mathbf{k}}^{(0)})f_{\mathbf{k}'}^{(0)}$
in terms of the difference $(f_{\mathbf{k}'}^{(0)}-f_{\mathbf{k}}^{(0)})$.
For that we use the relation \begin{equation}
(1-f_{\mathbf{k}}^{(0)})f_{\mathbf{k}'}^{(0)}=\frac{f_{\mathbf{k}'}^{(0)}-f_{\mathbf{k}}^{(0)}}{1-\exp[(\varepsilon_{\mathbf{k}'}-\varepsilon_{\mathbf{k}})/k_{B}T]},\label{eq:rel}\end{equation}
which is easily verified by direct calculation. For the case of Eq.~\eqref{eq:var1b},
where $\varepsilon_{\mathbf{k}'}-\varepsilon_{\mathbf{k}}=\hbar\omega_{\mathbf{q}}^{F}+\hbar\omega_{\mathbf{q}'}^{F}$
holds, we have\begin{equation}
\frac{1}{1-\exp[(\varepsilon_{\mathbf{k}'}-\varepsilon_{\mathbf{k}})/k_{B}T]}=-\frac{n_{\mathbf{q}}n_{\mathbf{q}'}}{1+n_{\mathbf{q}}+n_{\mathbf{q}'}},\end{equation}
while in the case of Eq.~\eqref{eq:var2b}, where $\varepsilon_{\mathbf{k}'}-\varepsilon_{\mathbf{k}}=\hbar\omega_{\mathbf{q}}^{F}-\hbar\omega_{\mathbf{q}'}^{F}$,
we get\begin{equation}
\frac{1}{1-\exp[(\varepsilon_{\mathbf{k}'}-\varepsilon_{\mathbf{k}})/k_{B}T]}=\frac{n_{\mathbf{q}}(1+n_{\mathbf{q}'})}{n_{\mathbf{q}}-n_{\mathbf{q}'}}.\end{equation}
The variations in Eqs.~\eqref{eq:var1a} and~\eqref{eq:var2a} may
then be cast in the form\begin{multline}
\delta\left[f_{\mathbf{k}'}(1-f_{\mathbf{k}})(n_{\mathbf{q}}+1)(n_{\mathbf{q}'}+1)-f_{\mathbf{k}}(1-f_{\mathbf{k}'})n_{\mathbf{q}}n_{\mathbf{q}'}\right]=\\
-(f_{\mathbf{k}'}^{(0)}-f_{\mathbf{k}}^{(0)})\frac{\partial n_{\mathbf{q}}^{(0)}}{\partial(\hbar\omega_{\mathbf{q}}^{F})}\frac{\partial n_{\mathbf{q}'}^{(0)}}{\partial(\hbar\omega_{\mathbf{q}'}^{F})}\frac{k_{B}T\left(\varphi_{\mathbf{k}'}-\varphi_{\mathbf{k}}-\chi_{\mathbf{q}}-\chi_{\mathbf{q}'}\right)}{1+n_{\mathbf{q}}+n_{\mathbf{q}'}},\label{eq:var1c}\end{multline}
and\begin{multline}
\delta\left[f_{\mathbf{k}'}(1-f_{\mathbf{k}})(n_{\mathbf{q}}+1)n_{\mathbf{q}'}-f_{\mathbf{k}}(1-f_{\mathbf{k}'})n_{\mathbf{q}}(n_{\mathbf{q}'}+1)\right]=\\
(f_{\mathbf{k}'}^{(0)}-f_{\mathbf{k}}^{(0)})\frac{\partial n_{\mathbf{q}}^{(0)}}{\partial(\hbar\omega_{\mathbf{q}}^{F})}\frac{\partial n_{\mathbf{q}'}^{(0)}}{\partial(\hbar\omega_{\mathbf{q}'}^{F})}\frac{k_{B}T\left(\varphi_{\mathbf{k}'}+\chi_{\mathbf{q}'}-\varphi_{\mathbf{k}}-\chi_{\mathbf{q}}\right)}{n_{\mathbf{q}}-n_{\mathbf{q}'}}.\label{eq:var2c}\end{multline}
Inserting the latter results into the variation of the two phonon
collision integral in Eq.~\eqref{eq:collInt2ph}, and recalling that
there are two other terms which can be obtained from Eqs.~\eqref{eq:var1c}
and~\eqref{eq:var2c} by multiplying a minus sign and substituting
$\mathbf{k}\rightarrow\mathbf{k}'$, we obtain\begin{widetext}\begin{eqnarray}
\delta\dot{f}_{\mathbf{k}}\bigr|_{\mathrm{scatt}} & = & -\frac{\pi}{\hbar}\sum_{\mathbf{k}'}\sum_{\mathbf{q},\mathbf{q}'}w_{F}(\mathbf{q},\mathbf{q}',\mathbf{k},\mathbf{k}')(f_{\mathbf{k}'}^{(0)}-f_{\mathbf{k}}^{(0)})\frac{\partial n_{\mathbf{q}}^{(0)}}{\partial(\hbar\omega_{\mathbf{q}}^{F})}\frac{\partial n_{\mathbf{q}'}^{(0)}}{\partial(\hbar\omega_{\mathbf{q}'}^{F})}\frac{k_{B}T}{1+n_{\mathbf{q}}+n_{\mathbf{q}'}}\left(\varphi_{\mathbf{k}'}-\varphi_{\mathbf{k}}-\chi_{\mathbf{q}}-\chi_{\mathbf{q}'}\right)\times\nonumber \\
 &  & \delta_{\mathbf{k},\mathbf{k}'-\mathbf{q}-\mathbf{q}'}\delta(\varepsilon_{\mathbf{k}}-\varepsilon_{\mathbf{k}'}+\hbar\omega_{\mathbf{q}}^{F}+\hbar\omega_{\mathbf{q}'}^{F})\nonumber \\
 & + & \frac{\pi}{\hbar}\sum_{\mathbf{k}'}\sum_{\mathbf{q},\mathbf{q}'}w_{F}(\mathbf{q},\mathbf{q}',\mathbf{k},\mathbf{k}')(f_{\mathbf{k}'}^{(0)}-f_{\mathbf{k}}^{(0)})\frac{\partial n_{\mathbf{q}}^{(0)}}{\partial(\hbar\omega_{\mathbf{q}}^{F})}\frac{\partial n_{\mathbf{q}'}^{(0)}}{\partial(\hbar\omega_{\mathbf{q}'}^{F})}\frac{k_{B}T}{n_{\mathbf{q}}-n_{\mathbf{q}'}}\left(\varphi_{\mathbf{k}'}+\chi_{\mathbf{q}'}-\varphi_{\mathbf{k}}-\chi_{\mathbf{q}}\right)\times\nonumber \\
 &  & \delta_{\mathbf{k},\mathbf{k}'-\mathbf{q}+\mathbf{q}'}\delta(\varepsilon_{\mathbf{k}}-\varepsilon_{\mathbf{k}'}+\hbar\omega_{\mathbf{q}}^{F}-\hbar\omega_{\mathbf{q}'}^{F})\nonumber \\
 & + & \frac{\pi}{\hbar}\sum_{\mathbf{k}'}\sum_{\mathbf{q},\mathbf{q}'}w_{F}(\mathbf{q},\mathbf{q}',\mathbf{k},\mathbf{k}')(f_{\mathbf{k}'}^{(0)}-f_{\mathbf{k}}^{(0)})\frac{\partial n_{\mathbf{q}}^{(0)}}{\partial(\hbar\omega_{\mathbf{q}}^{F})}\frac{\partial n_{\mathbf{q}'}^{(0)}}{\partial(\hbar\omega_{\mathbf{q}'}^{F})}\frac{k_{B}T}{1+n_{\mathbf{q}}+n_{\mathbf{q}'}}\left(\varphi_{\mathbf{k}'}+\chi_{\mathbf{q}}+\chi_{\mathbf{q}'}-\varphi_{\mathbf{k}}\right)\times\nonumber \\
 &  & \delta_{\mathbf{k},\mathbf{k}'+\mathbf{q}+\mathbf{q}'}\delta(\varepsilon_{\mathbf{k}}-\varepsilon_{\mathbf{k}'}-\hbar\omega_{\mathbf{q}}^{F}-\hbar\omega_{\mathbf{q}'}^{F})\nonumber \\
 & - & \frac{\pi}{\hbar}\sum_{\mathbf{k}'}\sum_{\mathbf{q},\mathbf{q}'}w_{F}(\mathbf{q},\mathbf{q}',\mathbf{k},\mathbf{k}')(f_{\mathbf{k}'}^{(0)}-f_{\mathbf{k}}^{(0)})\frac{\partial n_{\mathbf{q}}^{(0)}}{\partial(\hbar\omega_{\mathbf{q}}^{F})}\frac{\partial n_{\mathbf{q}'}^{(0)}}{\partial(\hbar\omega_{\mathbf{q}'}^{F})}\frac{k_{B}T}{n_{\mathbf{q}}-n_{\mathbf{q}'}}\left(\varphi_{\mathbf{k}'}+\chi_{\mathbf{q}}-\varphi_{\mathbf{k}}-\chi_{\mathbf{q}'}\right)\times\nonumber \\
 &  & \delta_{\mathbf{k},\mathbf{k}'+\mathbf{q}-\mathbf{q}'}\delta(\varepsilon_{\mathbf{k}}-\varepsilon_{\mathbf{k}'}-\hbar\omega_{\mathbf{q}}^{F}+\hbar\omega_{\mathbf{q}'}^{F}).\label{eq:linColIntTot2ph}\end{eqnarray}
\end{widetext}

Now we introduce the two typical approximations: consider phonons
to be in equilibrium, $\chi_{\mathbf{q}}\approx0$, so that $n_{\mathbf{q}}\approx n_{\mathbf{q}}^{(0)}$;
assume quasielastic scattering. From the latter we obtain $\delta(\varepsilon_{\mathbf{k}}-\varepsilon_{\mathbf{k}'}\pm\hbar\omega_{\mathbf{q}}^{F}\pm\hbar\omega_{\mathbf{q}'}^{F})\rightarrow\delta(\varepsilon_{\mathbf{k}}-\varepsilon_{\mathbf{k}'})$
and $f_{\mathbf{k}'}^{(0)}-f_{\mathbf{k}}^{(0)}\approx\pm\frac{\partial f_{\mathbf{k}}}{\partial\varepsilon_{\mathbf{k}}}\left(\hbar\omega_{\mathbf{q}}^{F}\pm\hbar\omega_{\mathbf{q}'}^{F}\right)$,
and the linearized collision integral then reads,\begin{multline}
\delta\dot{f}_{\mathbf{k}}\bigr|_{\mathrm{scatt}}=-\frac{2\pi}{\hbar^{2}}k_{B}T\frac{\partial f_{\mathbf{k}}}{\partial\varepsilon_{\mathbf{k}}}\sum_{\mathbf{k}'}\left(\varphi_{\mathbf{k}'}-\varphi_{\mathbf{k}}\right)\times\\
\sum_{\mathbf{q},\mathbf{q}'}w_{F}(\mathbf{q},\mathbf{q}',\mathbf{k},\mathbf{k}')\left(\frac{\omega_{\mathbf{q}}^{F}+\omega_{\mathbf{q}'}^{F}}{1+n_{\mathbf{q}}+n_{\mathbf{q}'}}-\frac{\omega_{\mathbf{q}}^{F}-\omega_{\mathbf{q}'}^{F}}{n_{\mathbf{q}}-n_{\mathbf{q}'}}\right)\times\\
\frac{\partial n_{\mathbf{q}}}{\partial\omega_{\mathbf{q}}^{F}}\frac{\partial n_{\mathbf{q}'}}{\partial\omega_{\mathbf{q}'}^{F}}\delta_{\mathbf{k},\mathbf{k}'+\mathbf{q}+\mathbf{q}'}\delta(\epsilon_{\mathbf{k}}-\epsilon_{\mathbf{k}'}),\label{eq:linColInt2ph}\end{multline}
where we have used the fact that $w_{F}(\mathbf{q},\mathbf{q}',\mathbf{k},\mathbf{k}')$
is invariant under changes $\mathbf{q}\rightarrow-\mathbf{q}$ and
$\mathbf{q}'\rightarrow-\mathbf{q}'$, and $\omega_{\mathbf{q}}^{F}=\omega_{-\mathbf{q}}^{F}$.
It can be written in the form of Eq.~\eqref{eq:linColIntQE}, with
$\mathcal{P}_{\mathbf{k},\mathbf{k}'}$ as given in Eq.~\eqref{eq:P2ph}.

\section{Calculating the resistivity}

\label{sec:APPres}

In this appendix we provide details regarding the calculation of the
$T-$dependent resistivity for bilayer graphene. The variational method
is used, with resistivity given by Eq.~\eqref{eq:resVarGraphene}.
Writing the numerator in Eq.~\eqref{eq:resVarGraphene} as in Eq.~\eqref{eq:current},
the resistivity can be cast in the form\begin{equation}
\varrho=\frac{\mathcal{V}\hbar^{2}}{8\pi^{2}e^{2}k_{F}^{4}}\int d\mathbf{k}d\mathbf{k}'\left(\mathbf{K}\cdot\mathbf{u}\right)^{2}\mathcal{P}_{\mathbf{k},\mathbf{k}'}\,.\label{eq:varResGraphApp}\end{equation}
The remaining task is the calculation of the integral on the right
hand side of Eq.~\eqref{eq:varResGraphApp}.

\subsection{Scattering by in-plane phonons}

\label{sub:APPresInPlane}

This case follows closely the derivation to the Bloch $T^{5}$ law
in 3-dimensional metals.\cite{zimanEP} Inserting Eq.~\eqref{eq:P1ph}
for $\mathcal{P}_{\mathbf{k},\mathbf{k}'}$ into Eq.~\eqref{eq:varResGraphApp}
we get\begin{multline}
\varrho=\frac{\mathcal{V}\hbar}{4\pi e^{2}k_{F}^{4}}\int d\mathbf{k}d\mathbf{k}'\left(\mathbf{K}\cdot\mathbf{u}\right)^{2}\times\\
\sum_{\nu}w_{\nu}(\mathbf{K},\mathbf{k},\mathbf{k}')\omega_{\mathbf{K}}^{\nu}\frac{\partial n_{\mathbf{K}}^{(0)}}{\partial\omega_{\mathbf{K}}^{\nu}}\frac{\partial f_{\mathbf{k}}^{(0)}}{\partial\epsilon_{\mathbf{k}}}\delta(\epsilon_{\mathbf{k}}-\epsilon_{\mathbf{k}'}),\label{eq:rho1phApp1}\end{multline}
where we have already performed the sum over $\mathbf{q}$. We can
simplify the integral above by integrating over $k$ and $k'$ noting
the presence of $\delta(\epsilon_{\mathbf{k}}-\epsilon_{\mathbf{k}'})$
and $\frac{\partial f_{\mathbf{k}}}{\partial\epsilon_{\mathbf{k}}}\approx-\delta(\epsilon_{F}-\epsilon_{\mathbf{k}})$.
The result reads,\begin{multline}
\varrho\approx-\frac{\mathcal{V}\hbar}{4\pi e^{2}k_{F}^{4}}\left(\frac{m}{\hbar^{2}}\right)^{2}\times\\
\int d\theta_{\mathbf{k}}d\theta_{\mathbf{k}'}\left(\mathbf{K}\cdot\mathbf{u}\right)^{2}\sum_{\nu}w_{\nu}(\mathbf{K},k_{F}\hat{e}_{\mathbf{k}},k_{F}\hat{e}_{\mathbf{k}'})\omega_{\mathbf{K}}^{\nu}\frac{\partial n_{\mathbf{K}}^{(0)}}{\partial\omega_{\mathbf{K}}^{\nu}},\label{eq:rho1phApp2}\end{multline}
with $\mathbf{K}=k_{F}(\hat{e}_{\mathbf{k}}-\hat{e}_{\mathbf{k}'})$.
In order to proceed with the calculation we have to specify the kernel
$w_{\nu}(\mathbf{K},k_{F}\hat{e}_{\mathbf{k}},k_{F}\hat{e}_{\mathbf{k}'})$
given in Eq.~\eqref{eq:wKernelLT2L}. Making use of the matrix elements
in Eq.~\eqref{eq:potential} we get,\begin{multline}
w_{\nu}(\mathbf{K},k_{F}\hat{e}_{\mathbf{k}},k_{F}\hat{e}_{\mathbf{k}'})\equiv w_{\nu}(K)=\frac{\left[D_{B}^{\nu}(K/k_{F})\right]^{2}\hbar^{3}k_{F}^{2}K^{2}}{2\mathcal{V}\rho v_{F}^{2}m^{2}\omega_{K}^{\nu}},\label{eq:wKernel1phApp}\end{multline}
with $D_{B}^{\nu}(x)$ as given in Eq.~\eqref{eq:couplingBilIn},
and where we have used the relation $K=2k_{F}\sin(\theta_{\mathbf{k},\mathbf{k}'}/2)$.
The kernel depends only on $\theta_{\mathbf{k},\mathbf{k}'}$, or
equivalently$K$ (the norm of $\mathbf{K}$), as is the case of the
rest of factors in the integrand of Eq.~\eqref{eq:rho1phApp2} but
for $\left(\mathbf{K}\cdot\mathbf{u}\right)^{2}$. The latter can
be written as $\left(\mathbf{K}\cdot\mathbf{u}\right)^{2}=K^{2}\cos^{2}\gamma$,
and the angular integration is then conveniently done by integrating
over $\gamma$ keeping $\theta_{\mathbf{k},\mathbf{k}'}=\theta_{\mathbf{k}}-\theta_{\mathbf{k}'}\equiv\theta$
constant, and integrate over $\theta$ afterward, or equivalently
$K$. Using $d\theta=dK/\sqrt{k_{F}^{2}-K^{2}/4}$, the resistivity
becomes\begin{equation}
\varrho\approx-\frac{\mathcal{V}m^{2}}{2\hbar^{3}e^{2}k_{F}^{4}}\sum_{\nu}\int_{0}^{2k_{F}}dK\frac{K^{2}w_{\nu}(K)\omega_{K}^{\nu}}{\sqrt{k_{F}^{2}-K^{2}/4}}\frac{\partial n_{K}^{(0)}}{\partial\omega_{K}^{\nu}}.\label{eq:rho1phApp3}\end{equation}
Inserting Eq.~\eqref{eq:wKernel1phApp} for the kernel $w_{\nu}(K)$
into Eq.~\eqref{eq:rho1phApp3} we readily obtain Eq.~\eqref{eq:resIn}.
From that it is straightforward to calculate analytically the two
limiting cases, $T\ll T_{BG}$ and $T\gg T_{BG}$.\cite{zimanEP}

\subsection{Scattering by flexural phonons}

\label{sub:APPresFlex}

Inserting Eq.~\eqref{eq:P2ph} for $\mathcal{P}_{\mathbf{k},\mathbf{k}'}$
into Eq.~\eqref{eq:varResGraphApp} we get\begin{multline}
\varrho=-\frac{\mathcal{V}k_{B}T}{4\pi e^{2}k_{F}^{4}}\int d\mathbf{k}d\mathbf{k}'\left(\mathbf{K}\cdot\mathbf{u}\right)^{2}\times\\
\sum_{\mathbf{q}}w_{F}(\mathbf{q},\mathbf{K-\mathbf{q}},\mathbf{k},\mathbf{k}')\frac{\partial n_{\mathbf{q}}}{\partial\omega_{\mathbf{q}}^{F}}\frac{\partial n_{\mathbf{q}'}}{\partial\omega_{\mathbf{q}'}^{F}}\frac{\partial f_{\mathbf{k}}^{(0)}}{\partial\varepsilon_{\mathbf{k}}}\delta(\varepsilon_{\mathbf{k}}-\varepsilon_{\mathbf{k}'})\times\\
\left(\frac{\omega_{\mathbf{q}}^{F}+\omega_{\mathbf{K-\mathbf{q}}}^{F}}{1+n_{\mathbf{q}}+n_{\mathbf{K-\mathbf{q}}}}-\frac{\omega_{\mathbf{q}}^{F}-\omega_{\mathbf{K-\mathbf{q}}}^{F}}{n_{\mathbf{q}}-n_{\mathbf{K-\mathbf{q}}}}\right),\label{eq:rho2phApp1}\end{multline}
where we have already performed the sum over $\mathbf{q}'$. We can
simplify the integral above by integrating over $k$ and $k'$ noting
the presence of $\delta(\epsilon_{\mathbf{k}}-\epsilon_{\mathbf{k}'})$
and $\frac{\partial f_{\mathbf{k}}}{\partial\epsilon_{\mathbf{k}}}\approx-\delta(\epsilon_{F}-\epsilon_{\mathbf{k}})$.
The result reads,\begin{multline}
\varrho\approx\frac{\mathcal{V}k_{B}T}{4\pi e^{2}k_{F}^{4}}\left(\frac{m}{\hbar^{2}}\right)^{2}\int d\theta_{\mathbf{k}}d\theta_{\mathbf{k}'}\left(\mathbf{K}\cdot\mathbf{u}\right)^{2}\times\\
\sum_{\mathbf{q}}w_{F}(\mathbf{q},\mathbf{K-\mathbf{q}},k_{F}\hat{e}_{\mathbf{k}},k_{F}\hat{e}_{\mathbf{k}'})\frac{\partial n_{\mathbf{q}}}{\partial\omega_{\mathbf{q}}^{F}}\frac{\partial n_{\mathbf{K-\mathbf{q}}}}{\partial\omega_{\mathbf{K-\mathbf{q}}}^{F}}\\
\left(\frac{\omega_{\mathbf{q}}^{F}+\omega_{\mathbf{K-\mathbf{q}}}^{F}}{1+n_{\mathbf{q}}+n_{\mathbf{K-\mathbf{q}}}}-\frac{\omega_{\mathbf{q}}^{F}-\omega_{\mathbf{K-\mathbf{q}}}^{F}}{n_{\mathbf{q}}-n_{\mathbf{K-\mathbf{q}}}}\right).\label{eq:rho2phApp2}\end{multline}
The kernel $w_{F}(\mathbf{q},\mathbf{K-\mathbf{q}},k_{F}\hat{e}_{\mathbf{k}},k_{F}\hat{e}_{\mathbf{k}'})$
is given by Eq.~\eqref{eq:wKernelLT2L} with $V_{\mathbf{q}}^{\nu}\rightarrow V_{\mathbf{q},\mathbf{q}'}^{F}$
(see Appendix~\ref{sec:APPcollInt} for a derivation). Inserting
the matrix elements in Eq.~\eqref{eq:potential} it takes the explicit
form\begin{multline}
w_{F}(\mathbf{q},\mathbf{K-\mathbf{q}},k_{F}\hat{e}_{\mathbf{k}},k_{F}\hat{e}_{\mathbf{k}'})\equiv w_{F}(q,K,|\mathbf{K-\mathbf{q}|})\\
=\frac{\left[D_{B}^{F}(K/k_{F})\right]^{2}\hbar^{4}q^{2}k_{F}^{2}|\mathbf{K}-\mathbf{q}|^{2}}{2^{4}\mathcal{V}^{2}m^{2}\rho^{2}v_{F}^{2}\omega_{q}^{F}\omega_{|\mathbf{K}-\mathbf{q}|}^{F}},\label{eq:wKernel2phApp}\end{multline}
with $D_{B}^{F}(x)$ as given in Eq.~\eqref{eq:couplingBilFlex},
and where we have used the relation $K=2k_{F}\sin(\theta_{\mathbf{k},\mathbf{k}'}/2)$
and assumed $\omega_{\mathbf{q}}^{F}$ given by Eq.~\eqref{eq:flexDispStr4}.
In deriving Eq.~\eqref{eq:wKernel2phApp} we used $\cos^{2}(\phi-\phi')=[1+\cos(2\phi-2\phi')]/2$
and dropped the oscillatory part. The sum over $\mathbf{q}$ can be
replaced by an integral, $\sum_{\mathbf{q}}\rightarrow\frac{\mathcal{V}}{(2\pi)^{2}}\int qdqd\phi$,
and owing to the relation $Q^{2}\equiv|\mathbf{K}-\mathbf{q}|^{2}=K^{2}+q^{2}-2qK\cos\phi$
we can write the resistivity as\begin{multline}
\varrho\approx\frac{\mathcal{V}^{2}k_{B}T}{8\pi^{3}e^{2}k_{F}^{4}}\left(\frac{m}{\hbar^{2}}\right)^{2}\int d\theta_{\mathbf{k}}d\theta_{\mathbf{k}'}\left(\mathbf{K}\cdot\mathbf{u}\right)^{2}\int_{0}^{\infty}dq\, q\frac{\partial n_{q}}{\partial\omega_{q}^{F}}\times\\
\int_{|K-q|}^{|K+q|}dQ\frac{Qw_{F}(q,K,Q)}{\sqrt{q^{2}K^{2}-\Bigl(K^{2}+q^{2}-Q^{2}\Bigr)^{2}/4}}\frac{\partial n_{Q}}{\partial\omega_{Q}^{F}}\times\\
\left(\frac{\omega_{q}^{F}+\omega_{Q}^{F}}{1+n_{q}+n_{Q}}-\frac{\omega_{q}^{F}-\omega_{Q}^{F}}{n_{q}-n_{Q}}\right),\label{eq:rho2phApp3}\end{multline}
where we used $d\phi=dQ\, Q/\sqrt{q^{2}K^{2}-\Bigl(K^{2}+q^{2}-Q^{2}\Bigr)^{2}/4}$.
As in Sec.~\ref{sub:APPresInPlane}, the angular integration over
$\theta_{\mathbf{k}}$ and $\theta_{\mathbf{k}'}$ is conveniently
done by integrating over $\gamma$, with $\left(\mathbf{K}\cdot\mathbf{u}\right)^{2}=K^{2}\cos^{2}\gamma$,
keeping $\theta_{\mathbf{k},\mathbf{k}'}=\theta_{\mathbf{k}}-\theta_{\mathbf{k}'}\equiv\theta$
and $q$ and $|\mathbf{K-\mathbf{q}|}\equiv Q$ constant, and 
integrate over $\theta$ afterward, $q$ and $Q$. The resistivity may then
be written as\begin{multline}
\varrho\approx\frac{k_{B}T}{2^{6}\pi^{2}e^{2}\rho^{2}v_{F}^{2}k_{F}^{2}}\int_{0}^{2k_{F}}dK\frac{\left[D_{B}^{F}(K/k_{F})\right]^{2}K^{2}}{\sqrt{k_{F}^{2}-K^{2}/4}}\times\\
\int_{0}^{\infty}dq\,\frac{q^{3}}{\omega_{q}^{F}}\frac{\partial n_{q}}{\partial\omega_{q}^{F}}\int_{|K-q|}^{|K+q|}dQ\frac{Q^{3}}{\omega_{Q}^{F}\sqrt{q^{2}K^{2}-\Bigl(K^{2}+q^{2}-Q^{2}\Bigr)^{2}/4}}\times\\
\frac{\partial n_{Q}}{\partial\omega_{Q}^{F}}\left(\frac{\omega_{q}^{F}+\omega_{Q}^{F}}{1+n_{q}+n_{Q}}-\frac{\omega_{q}^{F}-\omega_{Q}^{F}}{n_{q}-n_{Q}}\right),\label{eq:rho2phApp4}\end{multline}
where $d\theta=dK/\sqrt{k_{F}^{2}-K^{2}/4}$ has been used, and we
used Eq.~\eqref{eq:wKernel2phApp} for the kernel.

\subsubsection{Non-strained flexural phonons}

In the absence of strain the FP dispersion reads $\omega_{q}^{F}=\alpha q^{2}$.
After rescaling momentum as $x\rightarrow\tilde{x}=x(\hbar\alpha/k_{B}T)^{1/2}$
we can rewrite the resistivity as,\begin{multline}
\varrho\approx\frac{(k_{B}T)^{2}}{2^{6}\pi^{2}\hbar e^{2}\rho^{2}v_{F}^{2}k_{F}^{2}\alpha^{4}}\int_{0}^{2\tilde{k}_{F}}d\tilde{K}\frac{[D_{B}^{F}(\tilde{K}/\tilde{k}_{F})]^{2}\tilde{K}^{2}}{\sqrt{\tilde{k}_{F}^{2}-\tilde{K}^{2}/4}}\times\\
\int_{0}^{\infty}d\tilde{q}\tilde{q}n_{\tilde{q}}(n_{\tilde{q}}+1)\int_{|\tilde{K}-\tilde{q}|}^{|\tilde{K}+\tilde{q}|}d\tilde{Q}\frac{\tilde{Q}n_{\tilde{Q}}(n_{\tilde{Q}}+1)}{\sqrt{\tilde{q}^{2}\tilde{K}^{2}-\Bigl(\tilde{K}^{2}+\tilde{q}^{2}-\tilde{Q}^{2}\Bigr)^{2}/4}}\times\\
\left(\frac{\tilde{q}^{2}+\tilde{Q}^{2}}{1+n_{\tilde{q}}+n_{\tilde{Q}}}-\frac{\tilde{q}^{2}-\tilde{Q}^{2}}{n_{\tilde{q}}-n_{\tilde{Q}}}\right).\label{eq:rho2phAppNs1}\end{multline}
The integral over $\tilde{Q}$ is infrared divergent, and is thus
dominated by the contribution $\tilde{K}\sim\tilde{q}$. Defining
the small quantity $\delta x=|\tilde{K}-\tilde{q}|$, and noting that
for $\tilde{Q}\ll1$ we have $n_{\tilde{Q}}\sim1/\tilde{Q}^{2}\gg1$,
it is possible to identify the dominant contribution in the $\tilde{Q}$
integral as,\begin{multline*}
\int_{|\tilde{K}-\tilde{q}|}^{|\tilde{K}+\tilde{q}|}d\tilde{Q}\,\frac{\tilde{Q}n_{\tilde{Q}}(n_{\tilde{Q}}+1)}{\sqrt{\tilde{q}^{2}\tilde{K}^{2}-\Bigl(\tilde{K}^{2}+\tilde{q}^{2}-\tilde{Q}^{2}\Bigr)^{2}/4}}\times\\
\left(\frac{\tilde{q}^{2}+\tilde{Q}^{2}}{1+n_{\tilde{q}}+n_{\tilde{Q}}}-\frac{\tilde{q}^{2}-\tilde{Q}^{2}}{n_{\tilde{q}}-n_{\tilde{Q}}}\right)\sim2\tilde{K}^{2}\int_{\delta x}^{2\tilde{K}}d\tilde{Q}\frac{n_{\tilde{Q}}+1}{\tilde{K}}\sim\frac{2\tilde{K}}{\delta x}.\end{multline*}
It is now obvious that the $\tilde{q}$ integral has a logarithmic
divergence for $\tilde{q}\sim\tilde{K}$. Note, however, that in the
present theory phonons have an infrared cutoff, so that $\min|\tilde{K}-\tilde{q}|=\tilde{q}_{c}$,
where $\tilde{q}_{c}\ll1$ is either due to strain or anharmonic effects.
The dominant contribution to the $\tilde{q}$ integral is then coming
from the maximum of $1/|\tilde{K}-\tilde{q}|$, from which we obtain
\[
2\tilde{K}\int_{0}^{\infty}d\tilde{q}\,\tilde{q}n_{\tilde{q}}(n_{\tilde{q}}+1)\frac{1}{|\tilde{q}-\tilde{K}|}\sim-2\pi\tilde{K}^{2}n_{\tilde{K}}(n_{\tilde{K}}+1)\ln(\tilde{q}_{c}).\]
The resistivity may finally be written as a simple integral over $\tilde{K}$,\begin{multline}
\varrho\approx\frac{(k_{B}T)^{2}}{2^{6}\pi\hbar e^{2}\rho^{2}v_{F}^{2}k_{F}^{2}\alpha^{4}}\ln\left(\frac{k_{B}T}{\hbar\alpha q_{c}^{2}}\right)\times\\
\int_{0}^{2\tilde{k}_{F}}d\tilde{K}\frac{[D(\tilde{K}/\tilde{k}_{F})]^{2}}{\sqrt{\tilde{k}_{F}^{2}-\tilde{K}^{2}/4}}\tilde{K}^{4}n_{\tilde{K}}(n_{\tilde{K}}+1),\label{eq:rho2phAppNs2}\end{multline}
form which Eq.~\eqref{eq:resNsFlex} is readily obtained.

\subsubsection{Strained flexural phonons}

\label{sub:APPresFlexS}

The flexural phonon dispersion in the isotropic approximation is $\omega_{q}^{F}\approx\sqrt{\alpha^{2}q^{4}+\bar{u}v_{L}^{2}q^{2}}$
{[}see Eq.~\eqref{eq:flexDispStr4}{]}. After rescaling momenta $x\rightarrow\tilde{x}=x\hbar v_{L}u^{1/2}/(k_{B}T)$
the resistivity in Eq.~\eqref{eq:rho2phApp4} takes the form given
in Eq.~\eqref{eq:resStrFlex}. The low $T$ regime is detailed in
the main text. Here we concentrate in the high $T$ regime, showing
in particular how to obtain Eq.~(\ref{eq:Ggamma}) for the integrals
over $\tilde{q}$ and $\tilde{Q}$ in Eq.~\eqref{eq:resStrFlex}.

We start by writing the $\tilde{Q}$ integral in Eq.~\eqref{eq:resStrFlex}
as\begin{multline}
\mathcal{I}(\gamma,\tilde{K},\tilde{q})\equiv\int_{|\tilde{K}-\tilde{q}|}^{|\tilde{K}+\tilde{q}|}d\tilde{Q}\frac{\tilde{Q}^{3}n_{\tilde{Q}}(n_{\tilde{Q}}+1)}{\sqrt{\tilde{q}^{2}\tilde{K}^{2}-\Bigl(\tilde{K}^{2}+\tilde{q}^{2}-\tilde{Q}^{2}\Bigr)^{2}/4}}\times\\
\frac{1}{\sqrt{\gamma^{2}\tilde{Q}^{4}+\tilde{Q}^{2}}}\left(\frac{\sqrt{\gamma^{2}\tilde{q}^{4}+\tilde{q}^{2}}+\sqrt{\gamma^{2}\tilde{Q}^{4}+\tilde{Q}^{2}}}{1+n_{\tilde{q}}+n_{\tilde{Q}}}-\right.\\
\left.-\frac{\sqrt{\gamma^{2}\tilde{q}^{4}+\tilde{q}^{2}}-\sqrt{\gamma^{2}\tilde{Q}^{4}+\tilde{Q}^{2}}}{n_{\tilde{q}}-n_{\tilde{Q}}}\right),\label{eq:QintApp1}\end{multline}
with $\gamma=\sqrt{2}\omega_{q_{T}}^{F}/\omega_{q^{*}}^{F}$ Having
in mind that high $T$ implies $\tilde{K}\ll1$, we consider the integration
in Eq.~\eqref{eq:QintApp1} in two limiting cases: when $\tilde{q}\lesssim\tilde{K}\ll1$
and for $\tilde{q}\gg\tilde{K}$. In the former case, since $\tilde{q}\ll1$
and $\tilde{Q}\ll1$ hold, we can linearize the dispersion relation
and approximate the Bose-Einstein distribution function by $n_{\tilde{q}}\approx1/\tilde{q}$
and $n_{\tilde{Q}}\approx1/\tilde{Q}$ (as discussed in the main text,
finite strain implies $k_{F}\ll q^{*}$, so that the linearization
of the flexural phonon dispersion can be taken when $\tilde{q}\lesssim\tilde{K}\ll1$
holds). The integral over $\tilde{Q}$ in Eq.~\eqref{eq:QintApp1}
may then be approximated by\begin{multline}
\mathcal{I}(\gamma,\tilde{K},\tilde{q})\approx\int_{|\tilde{K}-\tilde{q}|}^{|\tilde{K}+\tilde{q}|}d\tilde{Q}\,\frac{2\tilde{q}\tilde{Q}}{\sqrt{\tilde{q}^{2}\tilde{K}^{2}-\Bigl(\tilde{K}^{2}+\tilde{q}^{2}-\tilde{Q}^{2}\Bigr)^{2}/4}},\label{eq:QintApp2}\end{multline}
and the integral can be done as\begin{align}
\mathcal{I}(\gamma,\tilde{K},\tilde{q}) & \approx\int_{|\tilde{K}-\tilde{q}|}^{|\tilde{K}+\tilde{q}|}d\tilde{Q}\frac{4\tilde{q}\tilde{Q}}{\sqrt{Y(\tilde{q},\tilde{Q},\tilde{K})}}\nonumber \\
 & =2\tilde{q}\arctan\left[\frac{\tilde{q}^{2}+\tilde{K}^{2}-\tilde{Q}^{2}}{\sqrt{Y(\tilde{q},\tilde{Q},\tilde{K})}}\right]_{|\tilde{K}-\tilde{q}|}^{|\tilde{K}+\tilde{q}|}=2\pi\tilde{q}\equiv\mathcal{I}(\tilde{q}),\label{eq:QintApp3}\end{align}
where we have defined \[
Y(\tilde{q},\tilde{Q},\tilde{K})=-(\tilde{q}-\tilde{K}-\tilde{Q})(\tilde{q}-\tilde{K}+\tilde{Q})(\tilde{q}+\tilde{K}-\tilde{Q})(\tilde{q}+\tilde{K}+\tilde{Q}).\]
On the other hand, for $\tilde{q}\gg\tilde{K}$ the integration region
is concentrated around $\tilde{q}$. We may then write the integral
in Eq.~\eqref{eq:QintApp1} as a slowly varying function, which we
can take out of the integral, multiplied by an integral of the form
of that in Eq.~\eqref{eq:QintApp2}, \begin{multline}
\mathcal{I}(\gamma,\tilde{K},\tilde{q})\approx\frac{\tilde{q}^{2}n_{\tilde{q}}(n_{\tilde{q}}+1)}{\sqrt{\gamma^{2}\tilde{q}^{4}+\tilde{q}^{2}}}\left(\frac{2\sqrt{\gamma^{2}\tilde{q}^{4}+\tilde{q}^{2}}}{1+2n_{\tilde{q}}}+\frac{1}{n_{\tilde{q}}(n_{\tilde{q}}+1)}\right)\times\\
\int_{|\tilde{K}-\tilde{q}|}^{|\tilde{K}+\tilde{q}|}d\tilde{Q}\,\frac{\tilde{Q}}{\sqrt{\tilde{q}^{2}\tilde{K}^{2}-\Bigl(\tilde{K}^{2}+\tilde{q}^{2}-\tilde{Q}^{2}\Bigr)^{2}/4}}\approx\\
\frac{\pi\tilde{q}^{2}n_{\tilde{q}}(n_{\tilde{q}}+1)}{\sqrt{\gamma^{2}\tilde{q}^{4}+\tilde{q}^{2}}}\left(\frac{2\sqrt{\gamma^{2}\tilde{q}^{4}+\tilde{q}^{2}}}{1+2n_{\tilde{q}}}+\frac{1}{n_{\tilde{q}}(n_{\tilde{q}}+1)}\right)\equiv\mathcal{I}(\gamma,\tilde{q}).\label{eq:QintApp4}\end{multline}
Since for $\tilde{q}\lesssim\tilde{K}$ the later result reduces to
$2\pi\tilde{q}$, as in Eq.~\eqref{eq:QintApp3}, we can use $\mathcal{I}(\gamma,\tilde{q})$
in Eq.~\eqref{eq:QintApp4} to approximate the $\tilde{Q}$ integral,
Eq.~\eqref{eq:QintApp1}, in the full region $\tilde{q}\lesssim\tilde{K}\ll1$
to $\tilde{q}\gg\tilde{K}$. This has been tested numerically to be
a good approximation as long as $\tilde{K}\ll1$. The $\tilde{q}$
integral in Eq.~\eqref{eq:resStrFlex} may then be cast in the $\tilde{K}$
independent form given in Eq.~\eqref{eq:Ggamma}.

\section{Perturbative treatment of anharmonic effects}

\label{secapp:anharmonic}

\begin{figure}
\begin{centering}
\includegraphics[width=0.75\columnwidth]{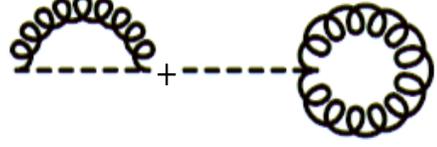}
\par\end{centering}

\caption{\label{fig:diagram}First order diagrams for $\Sigma\left(\mathbf{q}\right)$. The curly line represents the free correlator of flexural phonons, the dashed line corresponds to the 4-leg interaction vertex associated to $R_{ij,kl}\left(\mathbf{k}\right)$. The second diagram is $0$, since the $\mathbf{q}=0$ Fourier component of the transverse projector is integrated out during the in-plane modes Gaussian integration.}

\end{figure}

Recently it has been demonstrated how anharmonic effects in stiff membranes as graphene are highly suppressed by applying tension.\cite{RFZK11} In this Appendix we show how the infrared cutoff $q_{c}$ of our (harmonic) theory depends on the applied strain. This is consistent with the sample to sample differences of order unity reported in recent measurements of electron mobilities in doped suspended monolayer graphene.\cite{COK+10} As argued in Ref.~\onlinecite{COK+10}, comparing $q_c\approx0.1\textrm{\AA}^{-1}$,\cite{ZRF+10} (the infrared cutoff for the harmonic theory in the absence of strain), and $q^*$ [the momentum scale associated to the presence of strain, defined by Eq.~\eqref{eq:newEnScale}] gives $u\sim10^{-4}-10^{-3}$ as the strain involved in such kind of experiments. In order to estimate properly this number we have to know how $q_c$ is affected by strain. The discussion can be applied to both cases, monolayer or bilayer graphene.

In order to estimate anharmonic effects in the dispersion relation of FPs, since Eq.~\eqref{eq:fe} is quadratic in the in-plane displacements we can integrate them out to obtain the effective free-energy for the out-of-plane degree of freedom\cite{NelsonPeliti}\begin{equation}
\mathcal{F}^{eff}=\frac{1}{2}\kappa\int dx dy \left(\nabla^2h\right)^2+\frac{1}{2}\int dx dy R_{ij,kl}\partial^ih\partial^jh\partial^kh\partial^lh
\label{eq:effective_free_energy},
\end{equation}where the four-point-coupling fourth-order tensor can be written as $R_{ij,kl}=\frac{K_0}{4}P_{ij}^TP_{kl}^T$, the operator $P_{ij}^T$ is the transverse projector $P_{ij}^T=\left(\nabla^2\right)^{-1}\varepsilon_{ik}\varepsilon_{jl}\partial^k\partial^l$, and $K_0=\frac{4\mu\left(\mu+\lambda\right)}{2\mu+\lambda}$. In order to include the effect of strain, we add to Eq.~\eqref{eq:effective_free_energy} the simplest term which breaks rotational symmetry\begin{equation*}
\frac{1}{2}\gamma\int dxdy\left(\nabla h\right)^2
\label{eq:effective_free_energy_tension},
\end{equation*}
where $\gamma$ is a sample-dependent coefficient with units of tension which can be related with the strain of the sample. This approach follows the spirit of the effective isotropic dispersion relation of FPs introduced in Eq.~\eqref{eq:flexDispStr4}. If we add to Eq.~\eqref{eq:fe} the most general term to first order in the derivatives of the displacement fields which breaks rotational symmetry and then we integrate out the in-plane degrees of freedom, we obtain a new two-point vertex whose contribution to the renormalization of the bending rigidity is weak and can be neglected.\cite{RFZK11}. We are going to study the Fourier component of the height-height correlation function\begin{equation*}
G\left(\mathbf{q}\right)=\left\langle\left|h\left(\mathbf{q}\right)\right|^2\right\rangle=\frac{1}{Z}\int\mathcal{D}h\left(\mathbf{q}\right)\left|h\left(\mathbf{q}\right)\right|^2e^{S^{eff}\left[h\left(\mathbf{q}\right)\right]},
\end{equation*}where obviously $Z=\int\mathcal{D}h\left(\mathbf{q}\right)e^{S^{eff}}$ is the partition function of the system, the effective action is nothing but $S^{eff}\left[h\left(\mathbf{q}\right)\right]=-\beta F^{eff}\left[h\left(\mathbf{q}\right)\right]$, and the Fourier transformed effective free energy reads\begin{widetext}\begin{equation}
\mathcal{F}^{eff}\left[h\left(\mathbf{q}\right)\right]=\int\frac{d^2\mathbf{q}}{\left(2\pi\right)^2}\left(\frac{\kappa}{2}q^4+\frac{\gamma}{2}q^2\right)|h\left(\mathbf{q}
\right)|^2+\frac{1}{2}\int\prod_{i=1}^4\frac{d^2\mathbf{q}_i}{\left(2\pi\right)^8}q_1^iq_2^jq_3^kq_4^lR_{ij,kl}\left(\mathbf{q}_1+\mathbf{q}_2\right)\delta^{(2)}\left(\mathbf{q}_1+\mathbf{q}_2+\mathbf{q}_3+
\mathbf{q}_4\right),
\end{equation}\end{widetext}where $R_{ij,kl}\left(\mathbf{k}\right)=\frac{K_0}{4}P_{ij}^T\left(\mathbf{k}\right)P_{ij}^T\left(\mathbf{k}\right)$, and the Fourier transformed transverse projector reads $P_{ij}^T\left(\mathbf{k}\right)=k^{-2}\varepsilon_{ik}\varepsilon_{jl}k^kk^l=\delta_{ij}-\frac{k_ik_j}{k^2}$. It is important to note that the $\mathbf{q}=0$ Fourier component of the transverse projector is integrated out during the Gaussian integration of the in-plane modes.\cite{NelsonPeliti} In the harmonic approximation and in the absence of strain ($\gamma=0$), the (free) correlator is given by\begin{equation}
G^{(0)}\left(\mathbf{q}\right)=\frac{K_BT}{\kappa q^4}.
\end{equation}When we assume a quadratic dispersion relation we are taking this correlator as the proper one. This approximation is obviously affected by the presence of strain and anharmonic effects (also affected by strain), which renormalizes the bending rigidity $\kappa$. Then, we can write $G^{-1}\left(\mathbf{q}\right)\propto \kappa\left(\mathbf{q}\right)q^4$ and study the renormalization of $\kappa$ from the Dyson equation\begin{equation}
G^{-1}\left(\mathbf{q}\right)=\left(G^{(0)}\left(\mathbf{q}\right)\right)^{-1}+\Sigma\left(\mathbf{q}\right),\label{eq:Dyson}
\end{equation}where now the correlator in the harmonic theory, including the effect of strain, is given by\begin{equation}
G^{(0)}\left(\mathbf{q}\right)=\frac{K_BT}{\kappa q^4+\gamma q^2}.
\end{equation}In order to estimate the anharmonic effects we compute the first order diagrams for self-energy, showed in Fig.~\ref{fig:diagram}. Only the first diagram gives a non-zero contribution\begin{equation}
\Sigma^{(1)}\left(\mathbf{q}\right)=4\beta\int\frac{d^2\mathbf{k}}{\left(2\pi\right)^2}q^iq^jq^kq^lR_{ij,kl}\left(\mathbf{k}\right)G^{(0)}\left(\mathbf{q}-\mathbf{k}\right).
\end{equation}Replacing this result in equation Eq.~\eqref{eq:Dyson} we obtain\begin{equation}
\kappa\left(\mathbf{q}\right)=\kappa+\frac{\gamma}{q^2}+K_BTK_0\int\frac{d^2\mathbf{k}}{\left(2\pi\right)^2}\frac{\left(1-\frac{\left(\mathbf{q}\cdot\mathbf{k}\right)^2}
{q^2k^2}\right)^2}{\kappa|\mathbf{q}-\mathbf{k}|^4+\gamma|\mathbf{q}-\mathbf{k}|^2}.
\label{eq:kappa_eff}\end{equation}Follow the Ginzburg criterion,\cite{FLK07} we estimate the cutoff of the theory nothing but comparing each correcting term of Eq.~\eqref{eq:kappa_eff} with the bare value of $\kappa$. As we have already mentioned, there are two different cutoffs, the one given by strain in the harmonic approximation, and the other one associated to anharmonic effects. The first one is given by the second term of Eq.~\eqref{eq:kappa_eff}, and it is nothing but $q^*$,
\begin{equation}
q^*=\sqrt{\frac{\gamma}{\kappa}}.
\end{equation}Identifying this result with the momentum scale defined by Eq.~\eqref{eq:newEnScale}, we deduce the relation between $\gamma$ and $u$: $\gamma=\left(\lambda+2\mu\right)u$.

\begin{figure}
\begin{centering}
\includegraphics[width=0.95\columnwidth]{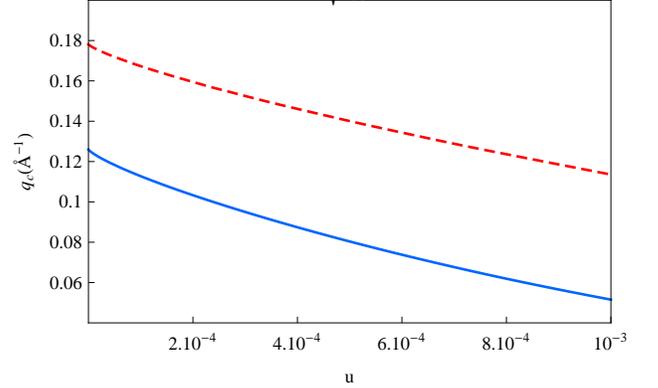}
\par\end{centering}

\caption{\label{fig:cutoff}The infrared cutoff $q_c$ as a function of the applied strain $u$ for monolayer (in red dashed line) and bilayer graphene (in blue). In both cases $T=300$ K.}

\end{figure}

The cutoff of the harmonic theory is also affected by strain. Following the same criterion, its value comes from the solution to the equation\begin{widetext}\begin{equation}
\kappa=\frac{K_BTK_0}{\left(2\pi\right)^2\kappa}\int_0^{2\pi} d\theta\int_0^{\infty}dk\frac{k\sin^4\left(\theta\right)}{\left(k^2+q^2-2kq\cos\left(\theta\right)\right)^2+\left(q^*\right)^2\left(k^2+q^2-2kq
\cos\left(\theta\right)\right)}.
\end{equation}\end{widetext}In the absence of strain, we have $q_c=\sqrt{\frac{3K_BTK_0}{16\pi\kappa^2}}$, which gives at $T=300$ K the value $q_c=0.178\textrm{\AA}^{-1}$ in the case of monolayer, and $q_c=0.126\textrm{\AA}^{-1}$ in the case of bilayer graphene. Its dependence on the applied strain is shown in Fig.~\ref{fig:cutoff}. It is clear that $q_c$ decreases as the strain increases, so the unavoidable little strain present in real samples increases the validity of the harmonic approximation.

%-----------------------------------------------------------------------------%
%-----------------------------------------------------------------------------%

%%%%%%%%%%%%%%%%%%%%%%%%%%%%%%%%%%%%%%%%%%%%%%%%%%%%%%%%%%%%%%%%%%%%%%%%%%%%%%%
\bibliographystyle{apsrev}

\begin{thebibliography}{41}
\bibitem[{foo()}]{foot1}
\bibinfo{note}{The fact that scattering is elastic and the Fermi surface
  isotropic makes unimportant factors depending only on $k$, the absolute value
  of momentum}.
\end{thebibliography}

\end{document}